\documentclass[twocolumn]{aastex63}

\usepackage{amsmath,amstext}
\usepackage{multirow}
\usepackage{rotating}
\usepackage{moresize}

\received{October 3, 2019}
\revised{May 31, 2020}
\submitjournal{AJ}

\shorttitle{BINOCS: Photometric Masses of Binary Star Systems in Clusters}
\shortauthors{Thompson et al.}

\begin{document}

\title{The Binary INformation from Open Clusters Using SEDs (BINOCS) Project: \\
Reliable Photometric Mass Determinations of Binary Star Systems in Clusters}

\correspondingauthor{Peter Frinchaboy}
\email{p.frinchaboy@tcu.edu}

\author{Benjamin Thompson}
\affiliation{Department of Physics \& Astronomy, Texas Christian University,
TCU Box 298840, \\Fort Worth, TX 76129, USA (p.frinchaboy, t.spoo,  j.donor@tcu.edu)}
\affiliation{GitHub, Inc.}

\author[0000-0002-0740-8346]{Peter M. Frinchaboy}
\affiliation{Department of Physics \& Astronomy, Texas Christian University,
TCU Box 298840, \\Fort Worth, TX 76129, USA (p.frinchaboy, t.spoo,  j.donor@tcu.edu)}

\author{Taylor Spoo}
\affiliation{Department of Physics \& Astronomy, Texas Christian University,
TCU Box 298840, \\Fort Worth, TX 76129, USA (p.frinchaboy, t.spoo,  j.donor@tcu.edu)}

\author{John Donor}
\affiliation{Department of Physics \& Astronomy, Texas Christian University,
TCU Box 298840, \\Fort Worth, TX 76129, USA (p.frinchaboy, t.spoo,  j.donor@tcu.edu)}

\begin{abstract}

We introduce a new binary detection technique, \textsc{Binary INformation from Open Clusters using SEDs (binocs)}, which we show is able to determine reliable stellar multiplicity and masses over a much larger mass range than current approaches.
This new technique determines accurate component masses of binary and single systems of the open clusters main sequence by comparing observed magnitudes from multiple photometric filters to synthetic star spectral energy distributions (SEDs) allowing systematically probing the binary population for {\em low mass} stars in clusters for 8 well-studied open clusters.
We provide new deep, infrared photometric catalogs ($1.2 - 8.0$ microns) for the key open clusters NGC 1960 (M36), NGC 2099 (M37), NGC 2420, and NGC2682 (M67), using observation from NOAO/NEWFIRM and {\em Spitzer}/IRAC.
Using these deep mulit-wavelength catalogs, the \textsc{binocs} method is applied to these clusters to determine accurate component masses for {\em unresolved} cluster binaries.  We explore binary fractions as a function of cluster age, Galactic location and metallicity.

\end{abstract}

\keywords{open clusters: general --- open clusters: individual (\objectname[M 37]{NGC 2099}, \objectname[M 67]{NGC 2682})}

\section{Introduction}

Binary stars have long been recognized to have an effect on stellar evolution, allowing the formation of non-standard stars, such as blue stragglers. While these stars may be formed in the field, their abundance in star clusters leads us to explore the effects of environment on the frequency of these stars. Additionally, it has long been recognized that the frequency of binaries will also affect the long term dynamics and stellar distributions within a cluster.

In order to more fully understand these effects on stellar and cluster evolution, the number of binaries within a cluster must be accurately determined, as well as other properties about the binary population. Having this data will allow correlations to be made between properties of the binary sample and dynamical traits of the cluster, as well as the frequency of anomalous stars. Correctly determining the number and composition of binary systems will allow deeper understanding of the physical processes behind stellar and cluster evolution.

Internal process can result in stars being ejected from the cluster due to gravitational interaction with other member stars. When a less-massive star gravitationally interacts with a more massive one, it may pick up enough energy to be accelerated beyond the escape velocity of the cluster. Binary systems may amplify this process by contributing part of their orbital energy to interactions, which is usually greater than the kinetic energy of the system moving through the cluster, which is fairly easy in poorly bound, low-mass, low-dispersion systems like open clusters.

Due to the vast timescale over which clusters evolve, stellar ejection cannot be studied observationally. Cluster ejection is usually studied via detailed N-Body simulations \citep{2001MNRAS.323..630H, 2005MNRAS.363..293H}, which can give a detailed description of what stars were ejected, when they were ejected, and how fast they were moving at the time of ejection. Each of these parameters dictate how the field population of the galaxy may have been built up by open cluster dissolution.

N-Body body simulations have already been run to analyze the binary population's effect on escaping stars, discovering only a slight difference when varying the cluster binary percentage from 0 to 70\% \citep{2013MNRAS.434.2509M}. These studies have made assumptions about the primordial binary population, however, such as an even mass-ratio (ratio of larger to smaller star mass) distribution, which may not be the case in reality.

Measuring the cluster binary fraction and mass-ratio distribution (as a function of primary mass) for real clusters will go a long way in calibrating these N-Body simulations. Analyzing open clusters with various ages will also allow N-Body simulations to check intermediate steps against these ``benchmarks,'' further improving their accuracy and predicting power.

It has also long been established that members of star clusters experience mass segregation. Binary systems, on average are more massive than single stars, have generally been thought to become more centrally concentrated than singles due to the same mechanism. A decreasing binary fraction with radius, indicative of mass segregation, has been observationally confirmed for several open and globular clusters \citep[e.g.,][]{2012AJ....144...54G, 2012AA...540A..16M}. Similar analyses have been conducted on the young ($15-30$ Myr), massive cluster NGC 1818, located in the Large Magellanic Cloud (LMC), producing conflicting results, which having more detailed clusters with binary characterization will allow us to fully explore.

While the advent of of high precision photometry, from {\em Kepler} and {\em TESS}, has opened up new studies for analyzing eclipsing binary systems, systematic methods for reliably and efficiently probing cluster binary populations for large numbers of clusters has remained elusive.

\section{Binary Detection}\label{sec:method}

\subsection{Previous Large-Scale Methods in Clusters}
Currently, binary systems within open clusters are detected using one of two methods. The first is \emph{two-band detection}, leveraged by both \citet{1998MNRAS.300..857E} and \citet{2013ApJ...765....4D}, which uses a cluster color-magnitude diagram (CMD) to quickly separate stars into singles and binaries. Stars lying far enough from the cluster single-star main sequence are classified as binaries, while all others are deemed single stars. While this method is quick and easy (only requiring imaging in two filters), it is plagued by degeneracies when attempting to determine accurate masses. In their paper, \citeauthor{2013ApJ...765....4D} admit that this method will only work inside a small mass range of NGC 1818. Outside of this region ``...the CMD is too steep to easily disentangle single from binary stars and blends. In addition, toward fainter magnitudes, photometric errors start to dominate any potential physical differences...'' To explore radial distributions of binary systems with a wide range of masses, two-band detection is not feasible.

The ideal method for this work is using radial velocity (RV) measurements to determine stellar multiplicity, as this approach will yield the most information about each system. There are, however, significant drawbacks: due to the number of stars in each cluster, and the velocity precision necessary to detect most binary systems, RV surveys can take hundreds of nights over many years, if not decades, to complete. Additionally, RV surveys cannot accurately measure cluster stars fainter than $V \sim 16$ without significant observing time on large telescopes, which removes a majority of cluster stars from the RV studies conducted so far.

With the growth of large-scale, space and ground-based photometric surveys \citep[e.g., Pan-STAARS, ESA {\em Gaia}, LSST, 2MASS, UKIDSS, VVV, {\em Roman} Space Telescope, {\em Spitzer}/GLIMPSE, {\em WISE}; ][]{panstarrs,gaia,lsst,2MASS_DR_allsky,ukidss,vvv,wfirst,glimpse,WISE_mission_description}, another photometry method would be ideal that is not cost-prohibitive in terms of telescope time, but will also yield accurate and detailed binary information over a wide range of stellar masses.

\vskip0.5in
\subsection{BINOCS}\label{sec:binocs_desc}
\begin{figure*} \centering
\epsscale{1.1}
\plotone{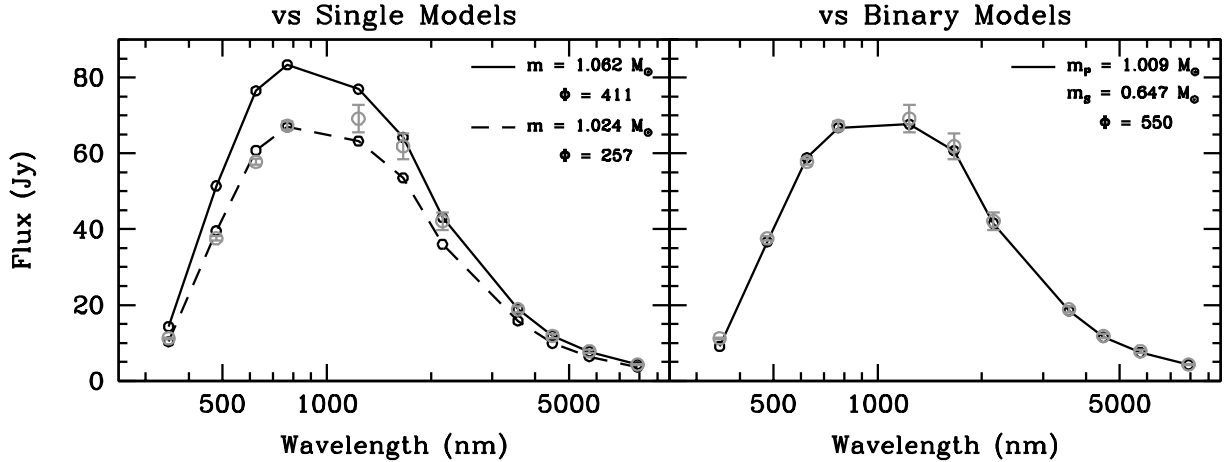} 
\caption{SED fitting of observed star in NGC 2682. \emph{(Left)} Comparison of observed fluxes (grey dots) to best-fit single star model SEDs. \emph{(Right)} Comparison of observed fluxes to best-fit single model. For each model, the fit's $\Phi$ value, defined in \S\ref{sec:code}, is shown. \label{fig:binocsEx}}
\end{figure*}

We introduce the new \textsc{Binary INformation from Open Clusters using Seds (binocs)} binary detection method, which determines accurate component masses of binary and single systems within open clusters by comparing observed magnitudes from multiple photometric filters to synthetic star spectral energy distributions (SEDs). An example of this method is shown in Figure \ref{fig:binocsEx}.

A star in NGC 2682 was observed in 11 bands ($ugriJHK_S$ [3.6][4.5][5.8][8.0])\footnote{IRAC [5.8] and [8.0] magnitudes provide little strength to the fit since they are on the Rayleigh- Jeans tail of the SED, and they are significantly more shallow than the [3.6] and [4.5] magnitudes, thus while they were used in testing the technique, we choose not to use them in the analysis presented here.}. When the observed magnitudes are compared against all single star model SEDs, two close-fitting models are detected, however neither fit the entire spectrum well (left panel of Figure \ref{fig:binocsEx}). While one model (with a mass of 1.062 M$_\odot$) fits accurately in the optical and $J$ band, it diverges for IR fluxes. The 1.024 M$_\odot$ model fits oppositely: overestimating optical fluxes, while accurately mapping the IR.

Next, the star is compared to binary model SEDs, where the best-fit is much more accurate (right panel of Figure \ref{fig:binocsEx}). A binary star in NGC 2682 with a primary mass of 1.009 M$_\odot$ and secondary mass of 0.647 M$_\odot$ fits within the observational uncertainties for 10 of the 11 observed bands. This star, while classified as a 'single' in a previous RV study \citep{1986AJ.....92.1364M}, is matched much more closely as a binary system using the \textsc{binocs} approach. Possible reasons for this mismatch will be further discussed in \S\ref{sec:rvcompare}.

A similar, but different Bayesian-based analysis has also recently been introduced by \citet{Cohen20}.

\subsection{BINOCS Code}\label{sec:code}

This \textsc{binocs} detection method is implemented through a publicly available code\footnote{\url{http://github.com/bathompso/binocs}}. The steps implemented by this code are described below.

First, the \textsc{binocs} code creates a library of synthetic cluster star SEDs using an isochrone, which lists stellar parameters ($T_\text{eff}$, $\log g$) and absolute magnitudes for a model star, given a cluster age, metallicity, reddening\footnote{The method works well over a range of reddening, but would be significantly adversely affected by {\em differential} reddening, which we plan to improve in future versions of the \textsc{binocs} code.}, and stellar mass. Isochrone sets often come in coarse mass grids, which hampers the \textsc{binocs} code's ability to compute accurate mass estimates. Stellar parameters are interpolated cubically in mass, generating new isochrone points in steps of 0.01 M$_\odot$. This interpolation only works along the main sequence, however, where mass increases monotonically. For evolved stars, at the turn-off or RGB, the original isochrone points are used.

Using the new mass-interpolated isochrone model, SEDs are created by computing up to 15 filter magnitudes ($UBVRIugrizJHK_S$[3.6][4.5]) for every possible combination of single synthetic stars in an isochrone. The isochrone absolute magnitudes are then adjusted to observed apparent magnitudes using the cluster's distance and reddening.

Next, each star in the cluster is compared to every possible model (binary and single) using:
\begin{equation} \label{eq:binocs}
\Phi = \sum_{\text{filters}} \frac{1}{\lvert m_{\text{star}} - m_{\text{model}} \rvert + \eta_{soft}}
\end{equation}

Here, $m_\text{star}$ is the observed magnitude of the star in a
particular band, while $m_\text{model}$ is the apparent magnitude of
the synthetic model star. $\eta_{soft}$ is a {\bf single global} softening parameter for which
we use $\eta_{soft} = 0.01$ (mag), the typical uncertainty of
the photometry.

If any of the sum elements is below a threshold value,  e.g., $\frac{1}{\lvert m_{\text{star}} - m_{\text{model}} \rvert + \eta_{soft}}  < 10$ (meaning the absolute difference in magnitudes is $> 0.09$), that element is declared to be ``distant'' and is not added to the sum.   The selection of this threshold values is explored in \S \ref{sec:filterthresh}. Only models with 3 good optical magnitudes ($UBVRIugriz$), 3 good near-IR magnitudes ($JHK_S$) and 2 good mid-IR  magnitudes ([3.6][4.5]) are considered. Figure \ref{fig:binocsEx} illustrates why such a requirement is necessary: binary SEDs differ from those of single stars only when compared across a large wavelength range. In a small region, the differences between a binary and single SED are negligible. Requiring a minimum number of good filter magnitudes across the entire wavelength range will ensure that all models accurately fit the entire SED, not just a single portion of it.

After discarding those models with too many ``distant'' magnitudes, all $\Phi$ values are normalized by the number of ``close'' magnitudes used in the sum. 
The model with the highest $\Phi$ is chosen as the best fit. If no models have enough ``close'' magnitudes, the star is marked as a non-member of the cluster.

After comparing each star to the full model library, it is also compared to only single models as a secondary check, using a much less stringent  ``close'' magnitudes cut --- each sum element must be $> 1$. The purpose of this comparison is two-fold: first is to be able to compare best-fitting single and binary models for illustrative purposes, as shown in Figure \ref{fig:binocsEx}. Secondly, some stars, while classified as binaries through the \textsc{binocs} method, are better classified as singles (these cuts will be explained in \S\ref{sec:minq}). If a star is forced to be classified as single, the parameters from this stage of fitting will be used to estimate its mass.

This fitting process is iterated 300 times, with each run randomly sampling magnitudes from a Gaussian distribution, with $\sigma$ equal to the photometric uncertainty. After all 300 resamples, the \textsc{binocs} code determines whether the star is a member or not. If a majority of the chosen best-fits denote that the star is a non-member, then that star is declared to be a non-member. Of the stars which are members, primary and secondary masses are determined by the median of all the best-fits. Uncertainties in the mass estimates are computed using the standard deviation of all best-fit masses. Similarly for the single-only fitting runs, the best-fit mass and uncertainty are the median and standard deviation of all results, respectively.

By using multiple filters over a large wavelength range (0.3 - 4.5 $\mu$m), individual photometric errors become less important than in the two-band detection method. This means that the \textsc{binocs} method can determine mass information for stars outside of the small mass window available for two-band detection. Additionally, only a small amount of telescope time, relative to RV surveys, is needed to capture cluster photometry across the optical to mid-IR (assuming access to the correct observing facilities). This allows for the detection of binaries in many clusters using a minimum of resources.

Additionally, we initially tested this technique using only  (0.3 - 2.2 $\mu$m), and found that the addition of the [3.6][4.5] significantly improved the resultant fitting.  Without utilizing the mid-IR filters resulting in significantly larger fitting errors and greater uncertainty in distinguishing binarity.

\S\ref{sec:data} will cover the photometric data used by the \textsc{binocs} routine in this work, while \S\ref{sec:isochrones} will discuss the underlying isochrone models used. \S\ref{sec:testing} will explore the assumptions in the method description above (number of good filters necessary, number of resamples, magnitude threshold level). \S\ref{sec:rvcompare} will compare \textsc{binocs} results to that of a previous radial velocity (RV) studies of NGC 2168 and 2682.

\section{Cluster Sample}
In total, 8 clusters were targeted for use in this work. The distribution of cluster parameters for our targeted sample is shown in Table \ref{tab:clusterParameters}. 

\begin{figure*} \centering
\plotone{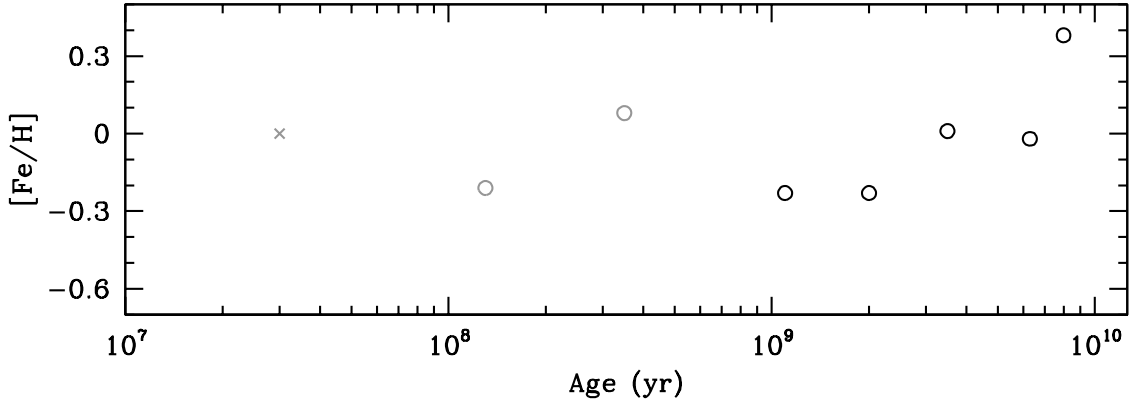}
\caption{Distribution of the eight targeted clusters in age and [Fe/H] \citep{2002AA...389..871D}. The 'X' represents the cluster NGC 1960 which does not have any published metallicity information, so here we assume solar metallicity.} 
\end{figure*}

\begin{table*} \centering \small
\caption{Adopted Cluster parameters for all clusters in dataset \citep{2002AA...389..871D}. \label{tab:clusterParameters}}
\begin{tabular}{lcccc}
\hline\hline \textbf{Cluster} & \textbf{Age (Gyr)} & \textbf{[Fe/H]} & \textbf{Dist(pc)} & \textbf{E(B-V)} \\ \hline 
NGC 188        & 6.30 & -0.02 & 1820 & 0.06 \\
NGC 1960 (M36) & 0.03 & \nodata  & 1320 & 0.22 \\
NGC 2099 (M37) & 0.35 & +0.08 & 1390 & 0.30 \\
NGC 2158       & 1.10 & -0.23 & 5080 & 0.36 \\
NGC 2168 (M35) & 0.13 & -0.21 &  870 & 0.20 \\
NGC 2420       & 2.00 & -0.23 & 2500 & 0.03 \\
NGC 2682 (M67) & 3.50 & +0.01 &  860 & 0.04 \\
NGC 6791       & 8.00 & +0.38 & 4170 & 0.15 \\\hline
\end{tabular}
\end{table*}

The cluster sample covers a large area of the parameter space: ages range from 25 Myr to 9 Gyr while [Fe/H] varies from $-0.23$ to $+0.38$ --- 40\% to 200\% the Iron content of the Sun. Exploiting this parameter range is critical in answering the posed science questions. In reference to science question 1, there are three clusters with ages $< 500$ Myr. Using \textsc{binocs} results from these three clusters, an understanding of the primordial cluster binary population can be conceived.

\section{Photometry Data} \label{sec:data}
The \textsc{binocs} method requires photometric data over a wide range of the spectrum (optical to mid-IR) to detect binaries effectively. Photometric data over this range was compiled from a variety of sources listed in Table \ref{tab:photsources}.

Table \ref{tab:photsources} summarizes the available data for use in this project, from the sources listed above, as well as from literature. 2MASS, WISE, and IRAC data are available for all clusters and are therefore not listed in Table \ref{tab:photsources}. Data sources in italics are not yet reduced, and not currently available for analysis.

\begin{table*} \centering \scriptsize
\caption{Photometry data fopr Clusters analyzed in this study\label{tab:photsources}}
\begin{tabular}{ l  c  c  c } \hline\hline
\textbf{Cluster} & \textbf{Visual Data} & \textbf{Near-IR Data} & \textbf{Membership Data} \\ \hline 
\multirow{2}{*}{NGC 188}	& \citet{1998AJ....116.1789V}	& \multirow{2}{*}{2MASS}	& \citet{2008AJ....135.2264G}	\\
							& \citet{2004PASP..116.1012S}	& 									& \citet{2003AJ....126.2922P}	\\\hline 
NGC 1960 (M36)				& MOSAIC						& NEWFIRM							& \citet{2000AA...357..471S}	\\ \hline 
NGC 2099 (M37)				& \citet{2008ApJ...675.1233H}	& NEWFIRM							& 								\\ \hline 
NGC 2158					& MOSAIC						& 2MASS				& 								\\ \hline 
NGC 2168 (M35)				& MOSAIC						& NEWFIRM							& \citet{Geller:2010hn}			\\ \hline 
NGC 2420					& \citet{2009ApJ...700..523A}	& NEWFIRM							&								\\ \hline 
\multirow{2}{*}{NGC 2682 (M67)}	& \citet{2009ApJ...700..523A}	& \multirow{2}{*}{NEWFIRM}			& \citet{Mathieu:1997tk}		\\
							& \citet{2008AA...484..609Y}	&									& \citet{2008AA...484..609Y}	\\\hline  
NGC 6791					& \citet{2009ApJ...700..523A}	& \citet{2005AJ....129..656C}		& 								\\ \hline
\end{tabular}
\end{table*}

Each of the cluster datasets in Table \ref{tab:photsources} have different levels of completeness, which will dictate which analysis projects the cluster can be included in. Clusters with complete photometry, although some may only have shallow 2MASS near-IR magnitudes, can have bulk binary population parameters determined, while complete deep photometry is necessary for the more detailed radial distribution analysis.

\subsection{Optical Photometry}
Many open clusters have been studied exhaustively using optical filters, including NGC 2099 and NGC 2682, and thus optical photometry for these clusters come from previously published sources.

NGC 2099 optical photometry is pulled from \citet{2008ApJ...675.1233H}, which used both short- and long-exposure images to provide $gri$ magnitudes for $10 \le r \le 23$.

NGC 2682 falls within the Sloan Digital Sky Survey \citep[SDSS;][]{2000AJ....120.1579Y} imaging region. The aperture photometry employed by SDSS caused problems for cluster photometry due to crowding in cluster core regions. \citet{2009ApJ...700..523A} reanalyzed the $ugriz$ SDSS images, extracting magnitudes using point-spread function (PSF) photometry, which can handle the dense cluster cores. Photometry in \citet{2009ApJ...700..523A} only covers two SDSS imaging regions near the core of NGC 2682, but does not touch further out regions. In these sparse outer regions, the original SDSS data release 7 \citep[DR7;][]{2009ApJS..182..543A} aperture photometry is accurate enough to be used.

Due to the length of exposure (54s) and telescope size (2.5-m), stars in the SDSS images begin to saturate above $r \sim 13$. Unfortunately, this removes almost all stars above the turn-off in NGC 2682. To fill in brighter stars that are not included in the SDSS catalog, $BVI$ photometry from \citet{2008AA...484..609Y} is used as a supplement. The combination of these two photometry sources provides nearly complete coverage of the cluster in the optical, from $V \sim 10$ to $g \sim 23$.

\subsubsection{MOSAIC:} The MOSAIC instrument \citep{2010SPIE.7735E..3AS}, outfitted with $UBVRI$ filters, contains an array of eight 2048-by-4096 pixel CCD chips to create a single 8192-by-8192 pixel image. While it has been attached previously to the 4-m telescope at Kitt Peak National Observatory (KPNO), the data used in this project is from the WIYN 0.9-m telescope at KPNO. With roughly a square degree field of view, the MOSAIC images will allow us to analyze the entire spatial extent of any cluster observed.

Images of several open clusters were obtained with MOSAIC over several nights in Feb 2000 (Sarajedini \& Kinemuchi, \emph{private communication}). $UBVI$ photometry was obtained on three clusters in the same set: M35, M36, M37. For all clusters, both short and long sequences of images were taken. Short images had exposure lengths of 25s, 8s, 5s, 5s in $UBVI$, respectively. Four images of the same exposure length were taken in each filter. Long sequence images, also four per filter, had 10 times the exposure length of the short set: 250s, 80s, 50s, 50s. Using both sequences together allows for photometry of the brightest and faintest stars within the cluster.

\begin{figure} \centering
\plotone{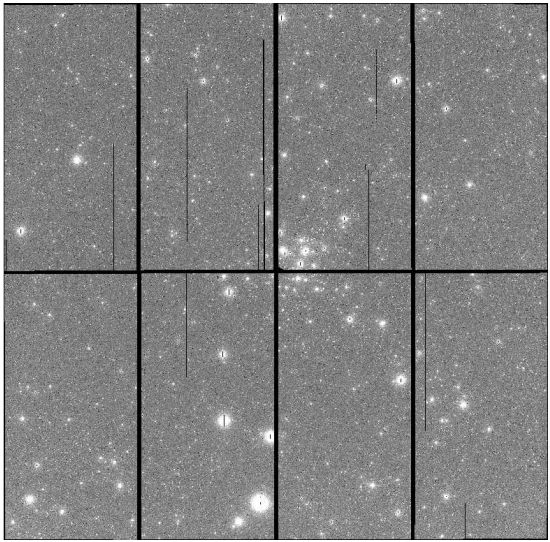}
\caption{Reduced MOSAIC 50s $I$-band image for NGC 1960.\label{fig:mosaic_image}}
\end{figure}

Two of the clusters have already been analyzed here: NGC 2168 in \citet{2014AJ....148...85T} and NGC 1960.  For our analysis, the four images in each filter were combined to form a higher signal-to-noise master image, and to provide a complete covering of the cluster. Note the wide gap between chips in the individual MOSAIC images, shown in Figure \ref{fig:mosaic_image}. Each of the four images per filter were \emph{dithered} (slightly offset) such that the combined image had no gaps in coverage. 

These master images were then split into the 8 individual chips on the MOSAIC image. This splitting was done to accommodate the DAOPHOT PSF photometry package, which has limits on image size. The individual 2k$\times$4k chips were the largest DAOPHOT could handle. In each chip (and for each master image), the process was the same. First, 400 candidate template stars were chosen to create a PSF. Next, the trimming process described in \citet{2014AJ....148...85T} was run, trimming the candidate list down to 250-300 template stars. Using this cleaned list, PSF parameters were determined, and then applied through ALLSTAR.

\begin{figure*} 
\plottwo{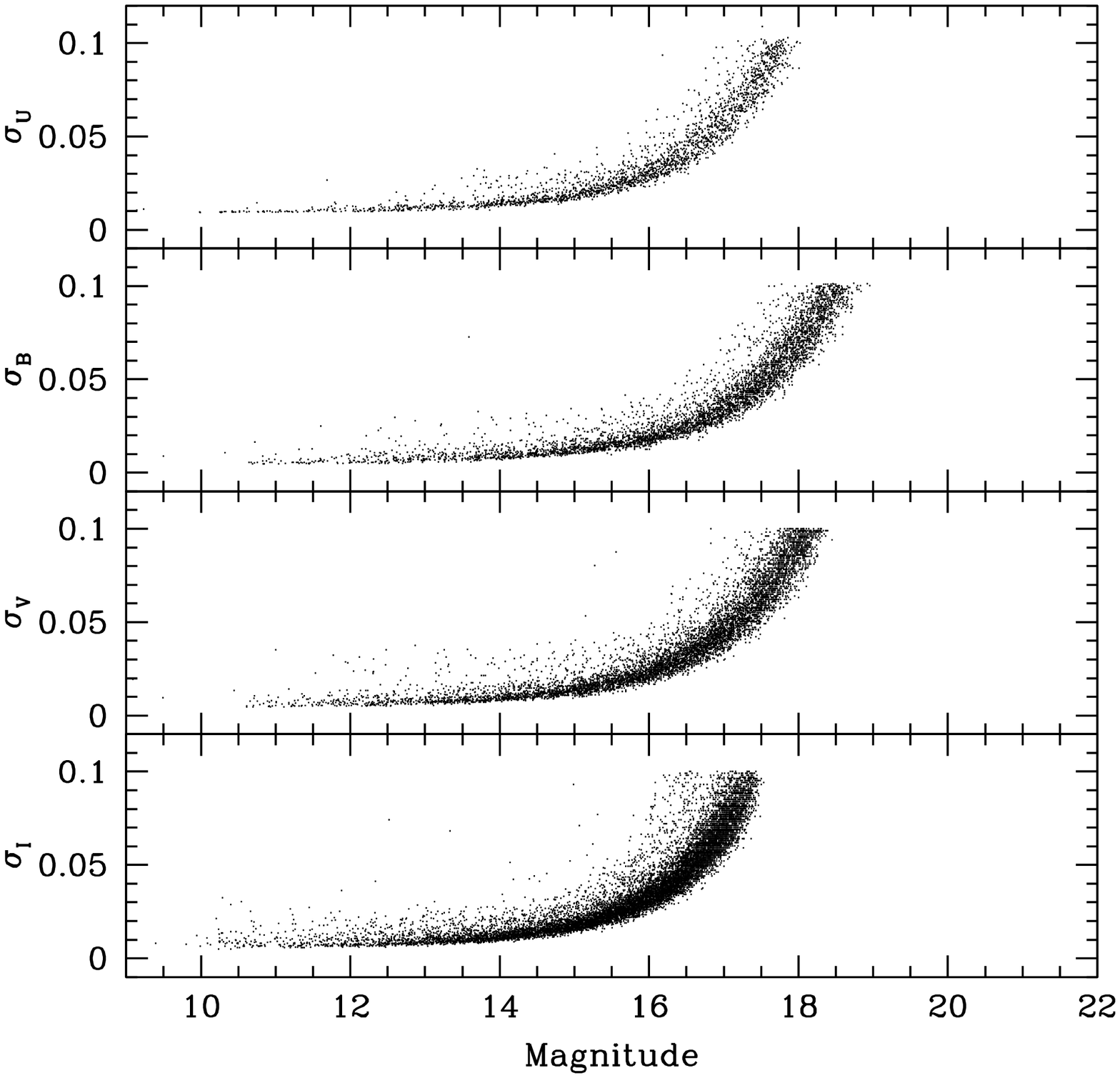}{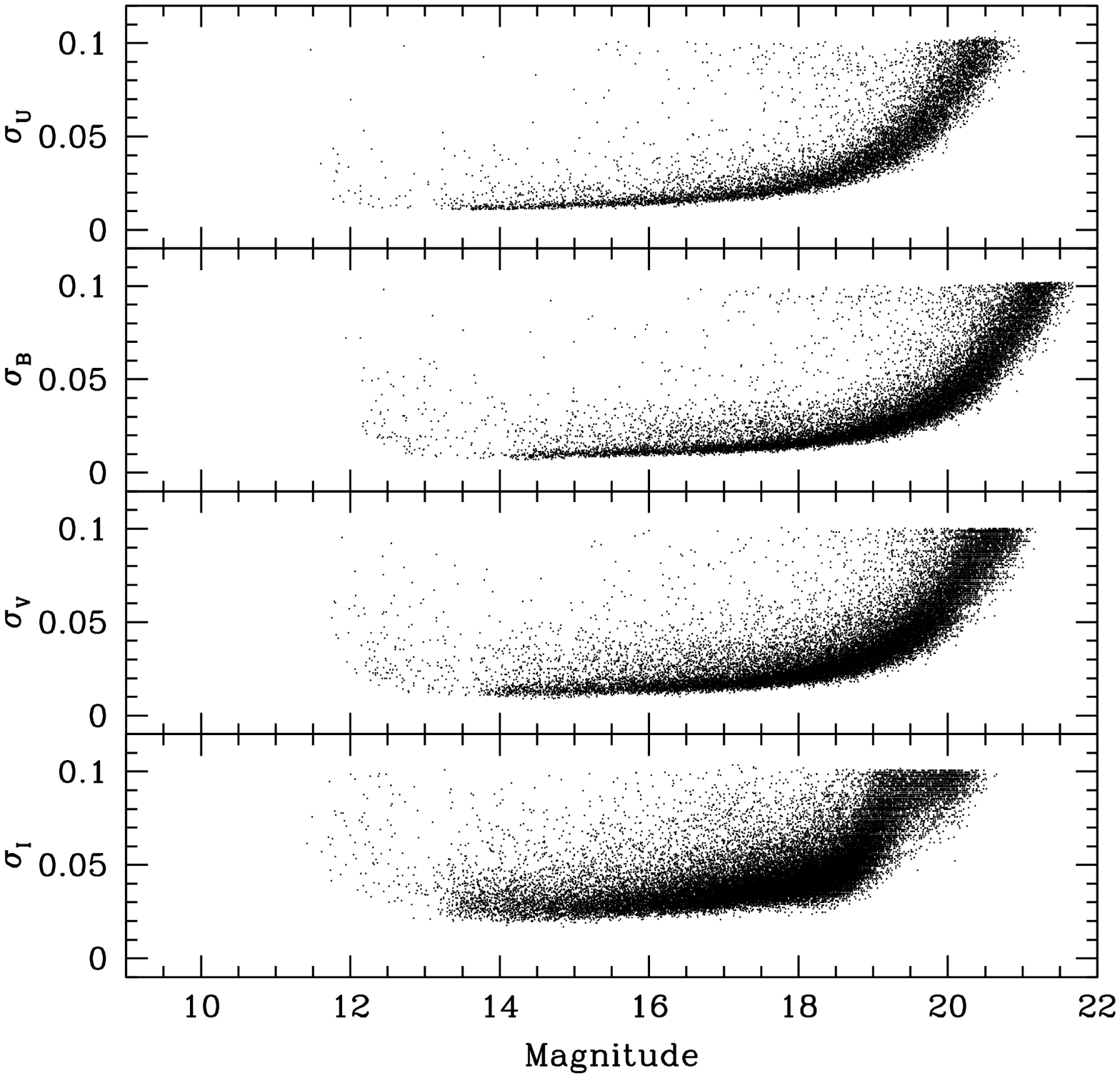}
\caption{MOSAIC photometric quality plots for NGC 1960 in $UBVI$. \emph{Left:} Short set of exposures. \emph{Right:} Long set of exposures. \label{fig:mosaic_quality}}
\end{figure*}

Photometric quality plots for the short and long sets are shown in Figure \ref{fig:mosaic_quality}. For reference, high quality photometry has uncertainties less than 0.05. The MOSAIC images provide this high quality data for $11 \le V \le 20$, covering nearly all of the stars within these clusters.

The ALLSTAR-derived magnitudes must be transformed to a standard system, in order to be comparable to other results. For calibration, photometry from the individual chips were re-combined to produce single photometry files for each master image, then matched to previously published ``standard'' photometry. For NGC 1960, the previously published $UBVI$ photometry from \citet{2006AJ....132.1669S} was used to transform the instrumental MOSAIC magnitudes to the standard system. 

Sources detected in the MOSAIC images were matched to the published photometry for each cluster, producing between $500-600$ matches for each filter. Using these common stars, the instrumental ALLSTAR magnitudes were transformed via the following equations: 

\begin{equation} \label{eq:Utrans}
	u = U + a_U + b_U \times (U-B)
\end{equation}
\begin{equation} \label{eq:Btrans}
	b = B + a_B + b_B \times (B-V)
\end{equation}
\begin{equation} \label{eq:Vtrans}
	v = V + a_V + b_V \times (B-V)
\end{equation}
\begin{equation} \label{eq:Itrans}
	i = I + a_I + b_I \times (V-I)
\end{equation}

\begin{table} \centering \small
\caption{Transformation coefficients for MOSAIC photometry.\label{tab:mosaic_coeffs}}
\begin{tabular}{ccccc} \hline\hline
Cluster & Filter & Length & a & b \\ \hline \hline
\multirow{8}{*}{NGC 1960} 
	& \multirow{2}{*}{$U$} & Short & $1.843 \pm 0.009$ & $0.008 \pm 0.011$ \\ & & Long & $-0.650 \pm 0.010$ & $-0.053 \pm 0.008$ \\ \cline{2-5}
	& \multirow{2}{*}{$B$} & Short & $1.191 \pm 0.004$ & $-0.105 \pm 0.005$ \\ & & Long & $-1.305 \pm 0.005$ & $-0.127 \pm 0.006$ \\ \cline{2-5}
	& \multirow{2}{*}{$V$} & Short & $1.536 \pm 0.003$ & $0.048 \pm 0.004$ \\ & & Long & $-0.928 \pm 0.005$ & $0.034 \pm 0.006$ \\ \cline{2-5}
	& \multirow{2}{*}{$I$} & Short & $1.993 \pm 0.004$ & $0.002 \pm 0.004$ \\ & & Long & $-0.562 \pm 0.011$ & $-0.000 \pm 0.011$ \\ \hline
\end{tabular}
\end{table}

\begin{figure} 
\plotone{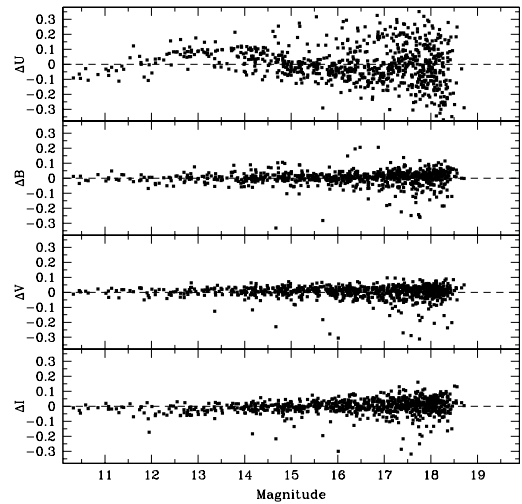}
\caption{Residuals from transformation to standard system for NGC 1960 MOSAIC photometry. \label{fig:mosaic_resid}}
\end{figure}

Here, lowercase filter letters indicate instrumental (ALLSTAR-derived) magnitudes, while uppercase filters are those of the standard photometry. The transformation coefficients for each cluster and filter are located in Table \ref{tab:mosaic_coeffs}. Transformations were done separately for the short and long exposure sequences. Residuals for these transformations are shown in Figure \ref{fig:mosaic_resid}. Once the instrumental magnitude were calibrated to the standard system, all photometry was combined into a single master catalog. \\

\subsection{$JHK_S$ Near-IR Photometry}
While Two-Micron All Sky Survey \citep[2MASS;][]{2006AJ....131.1163S} $JHK_S$ photometry is available for all open clusters, it is too shallow ($J \sim 16$) to provide photometry for low-mass members of the cluster. New $JHK_S$ near-IR photometry was obtained, by us, using the NEWFIRM instrument \citep{2004SPIE.5499...59H} on the Kitt Peak 4-m telescope. Images were taken on two nights in Feb 2008. Observation and reduction processes are the same as used in \citet{2014AJ....148...85T} from which we published the NGC 2168 data.

Observations were taken in ``4Q'' mode, which offsets the telescope in a pattern to align the center of the cluster on each of the four NEWFIRM detectors. This allows for larger area coverage than a single NEWFIRM field of view. To minimize errors in flat-fielding and negate cosmetic defects within the chips, the telescope was dithered between exposures for both clusters. Clusters have effective exposure times of 2400s in $J$ and $H$, and a total of 3600s in $K_S$.

All images were reduced (dark correction, flat fielding, sky subtraction) through the NEWFIRM pipeline \citep{2009ASPC..411..506S}. The reduced images were stacked into master frames for each filter. Photometry was carried out using the DAOPHOT II and ALLSTAR programs \citep{1987PASP...99..191S}, using a detection threshold of 3$\sigma$ in all filters. Initially, 2000 stars were chosen to determine the PSF for each stacked image. This list was then trimmed to remove stars which degraded the fit. First, stars near bad or saturated pixels were removed, so as to avoid PSF distortion by these outliers. Next, stars that were less than 4 full-width at half-maximum (FWHM) from another source were removed from the PSF list, ensuring that the PSF would not be contaminated by crowding. Finally, stars whose PSF $\chi^2$ fit values were 2$\sigma$ or more above the mean were removed. After trimming, approximately 500 and 800 stars remained for determining the PSF in NGC 2682 and NGC 2099, respectively.

The DAOPHOT-derived magnitudes were tied to the standard system by matching to 2MASS photometry. Only 2MASS sources with the highest photometric quality (`AAA') were used in the standard catalog. NGC 2682 frames matched approximately 700 stars between the NEWFIRM and 2MASS datasets, while NGC 2099 had almost 2000 overlapping sources. Using these matches, transformations were determined to the standard system for each cluster:
\begin{equation}\label{nfmjtran}
j = J + a_j (J-K_S) + b_j
\end{equation}
\begin{equation}\label{nfmhtran}
h = H + a_h (H-K_S) + b_h
\end{equation}
\begin{equation}\label{nfmktran}
k = K_S + a_k (J-K_S) + b_k
\end{equation}

In equations \eqref{nfmjtran}--\eqref{nfmktran}, lowercase letters denote DAOPHOT magnitudes, while uppercase letters denote 2MASS standard magnitudes. Transformation coefficients for each of the clusters is listed in table \ref{tab:nfmcoeffs}.

\begin{deluxetable*}{clll}
\tablecaption{NEWFIRM Transformation Coefficients \label{tab:nfmcoeffs}}
\tablecolumns{4}
\tablewidth{0pt}
\tablehead{
\colhead{Cluster} &
\colhead{$J$} &
\colhead{$H$} & 
\colhead{$K_S$}  
}
\startdata
\multirow{2}{*}{NGC 1960} & $a_j = -0.056 \pm 0.006$ & $a_h = -0.177 \pm 0.018$ & $a_k = +0.042 \pm 0.006$\\
                                                     & $b_j =  +2.441 \pm 0.004$ & $b_h = +2.620\pm0.003$ & $b_k = +3.063 \pm 0.004$\\\hline
\multirow{2}{*}{NGC 2099}  & $a_j = -0.121 \pm 0.008$ & $a_h = -0.354 \pm 0.016$ & $a_k = +0.112 \pm 0.009$ \\
                                                     & $b_j =  +2.434 \pm 0.004$ & $b_h =  +2.318 \pm 0.003$ & $b_k = +3.020 \pm 0.005$ \\\hline
\multirow{2}{*}{NGC 2168}  & $a_j = -0.099 \pm 0.005$ & $a_h = -0.296 \pm 0.012$ & $a_k = +0.093 \pm 0.007$\\
                                                     & $b_j = +2.397\pm0.003$ & $b_h = +2.297\pm0.002$ & $b_k = +3.030 \pm 0.005$\\\hline
\multirow{2}{*}{NGC 2420}           & $a_j = -0.036 \pm 0.008$ & $a_h = -0.234 \pm 0.020$ & $a_k = +0.130 \pm 0.011$\\
                                                     & $b_j = +2.752\pm0.005$ & $b_h = +2.739\pm0.003$ & $b_k = +3.179 \pm 0.006$\\\hline
\multirow{2}{*}{NGC 2682}  & $a_j = -0.100 \pm 0.010$ & $a_h = -0.250 \pm 0.021$ & $a_k = +0.113 \pm 0.014$ \\
                                                     & $b_j = +2.444 \pm 0.007$ & $b_h = +2.277 \pm 0.004$ & $b_k = +2.956 \pm 0.010$  
\enddata
\end{deluxetable*}

\begin{figure*} \centering
\epsscale{1.1}
\plottwo{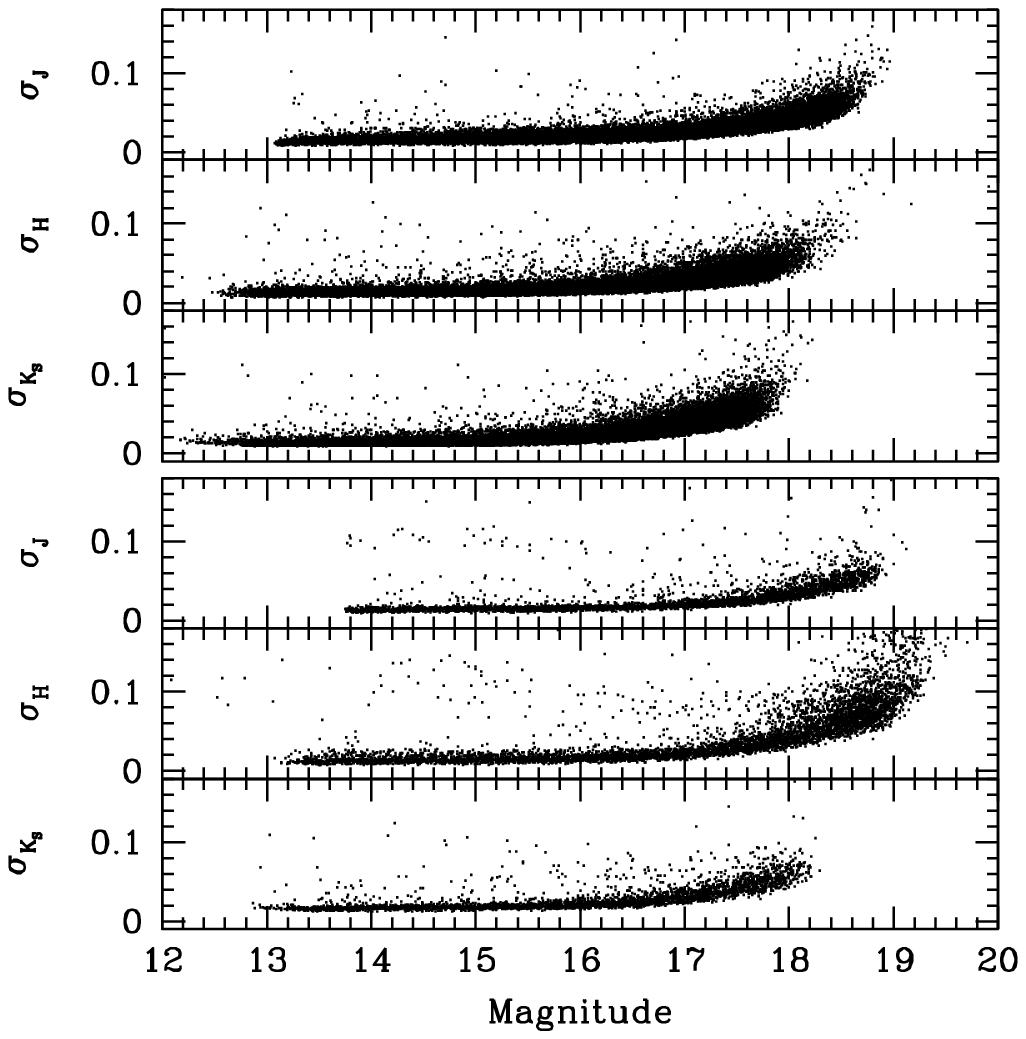}{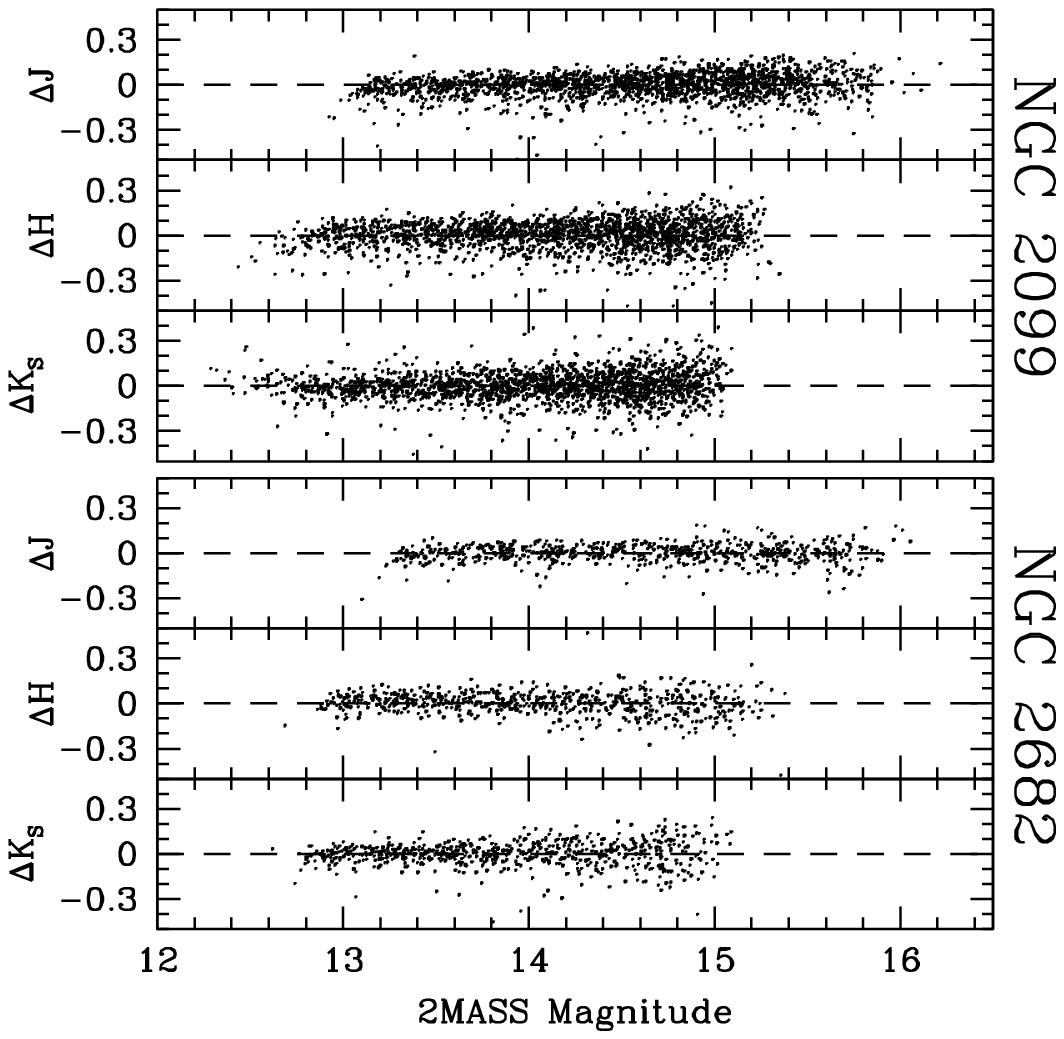}
\caption{\emph{Left:} NEWFIRM magnitude vs uncertainty for for NGC 2099 (top) \& NGC 2682 (bottom). \emph{Right:} Transformation residuals between NEWFIRM magnitudes and 2MASS.\label{fig:nfmphot}}
\end{figure*}

A plot of transformation residuals is shown in Figure \ref{fig:nfmphot}, along with measurement uncertainty as a function of magnitude. For almost all stars in the NEWFIRM images, magnitude uncertainties are $< 0.1$.

The NOAO/NEWFIRM photometry reaches a depth of approximately ($J$,$H$,$K_S$ = 18.6, 18.1, 17.8). 

\vskip0.7in
\subsection{Mid-IR Photometry}
Deep Mid-IR photometry ([3.6][4.5][5.8][8.0]) for NGC 2099 and NGC 2682 was gathered using the Infrared Array Camera \citep[IRAC;][]{2004ApJS..154...10F} on the Spitzer Space Telescope. NGC 2682 data was obtained as part of cycle 2 proposal 20710 (PI Skrutskie), and NGC 2099 data was obtained from cycle 3 proposal 30800 (PI Frinchaboy). The data were taken in High Dynamic Range (HDR) mode, allowing measurement of the brightest and faintest stars in the cluster simultaneously. The data were reduced and photometered using the GLIMPSE \citep[Galactic Legacy Infrared Mid-Plane Survey Extraordinare;][]{2003PASP..115..953B} pipeline, which was modified to handle the HDR data. The {\em Spitzer}/IRAC photometry reaches a depth of approximately ([3.6][4.5][5.8][8.0] = 18.0, 16.5, 14.6, 13.8 ). 
 Although within this work using the \textsc{binocs} method we use only the [3.6] and [4.5] bands, for completeness purposes, we provide this new photometry for all four Spitzer bands to the community in Table \ref{tab:NGC 2682stub}.

Supplemental Mid-IR photometry ($3-4.6\mu$m) is available from the Wide-field Infrared Survey Explorer \citep[WISE;][]{2010AJ....140.1868W} for the entire sky. WISE photometry was pulled for a 1$^\circ$ radius around the cluster, extending the spatial coverage of the mid-IR data. Unfortunately, WISE and IRAC use slightly different filters, and a correction must be applied to the WISE photometry in order to merge it with the deeper IRAC data.

WISE and IRAC photometry of NGC 2099 were compared, and transformation equations were found using more than 800 common sources. Transformations were limited to [3.4]$_{\text{WISE}} < 14$ and [4.6]$_{\text{WISE}} < 13.5$, beyond which the transformations become problematic. Residuals between IRAC and WISE photometry shows no correlation with color, and only small magnitude offsets.

\begin{figure} \centering
\epsscale{1.1}
\plotone{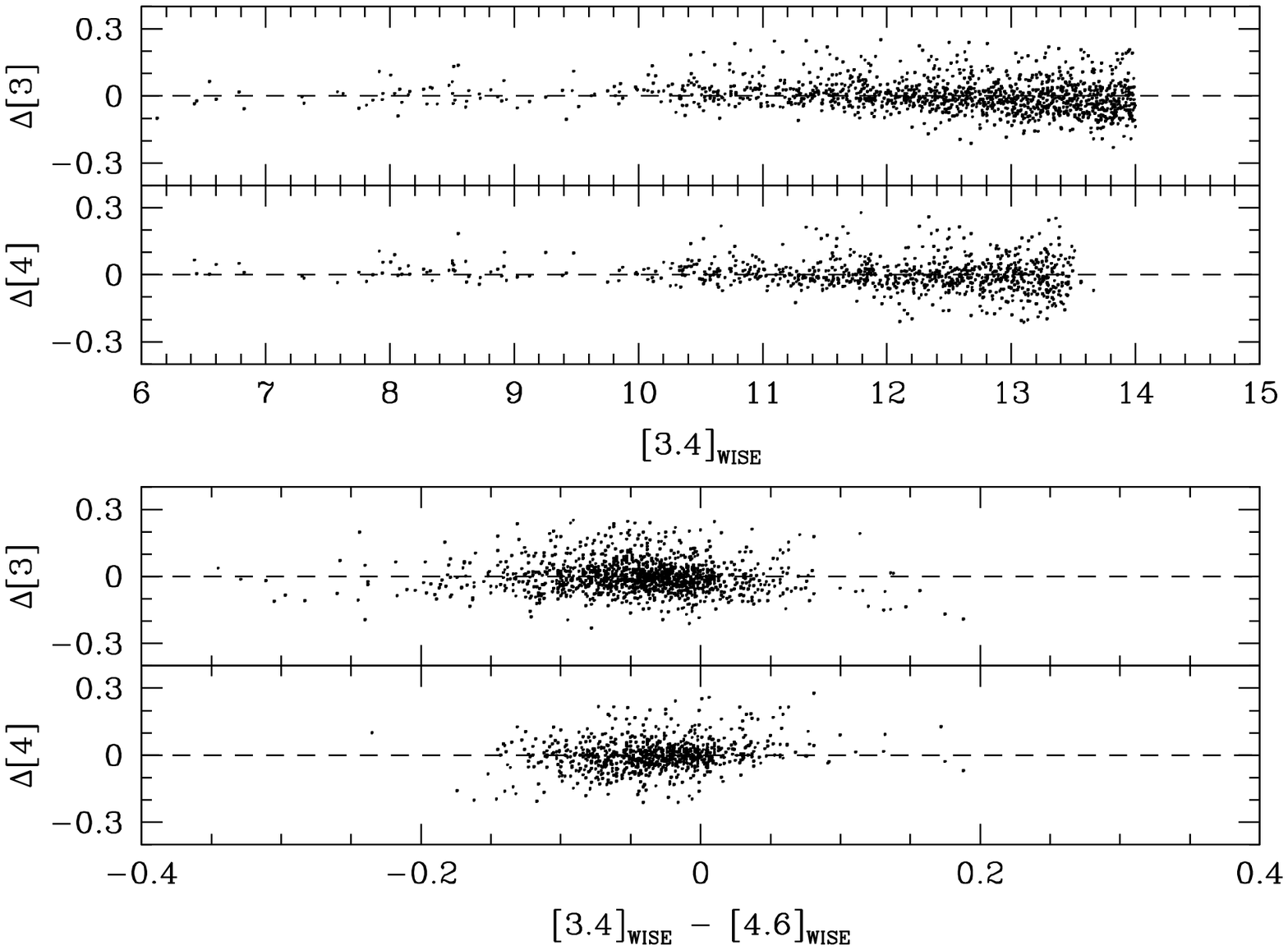}
\caption{Residuals from transformation between WISE and IRAC magnitudes for sources near NGC 2099.\label{fig:mirresid}}
\end{figure}

[3.4]$_{WISE}$ and [3.6]$_{IRAC}$ are interchangeable in the specified region, while [4.6]$_{WISE} = $ [4.5]$_{IRAC} + 0.03$. A plot of residuals for this transformation are shown in Figure \ref{fig:mirresid}. Using these simple transformation equations, WISE photometry was merged with the IRAC sources.

\subsection{Merged Dataset}
\begin{figure*} \centering
\epsscale{1.1}
\plotone{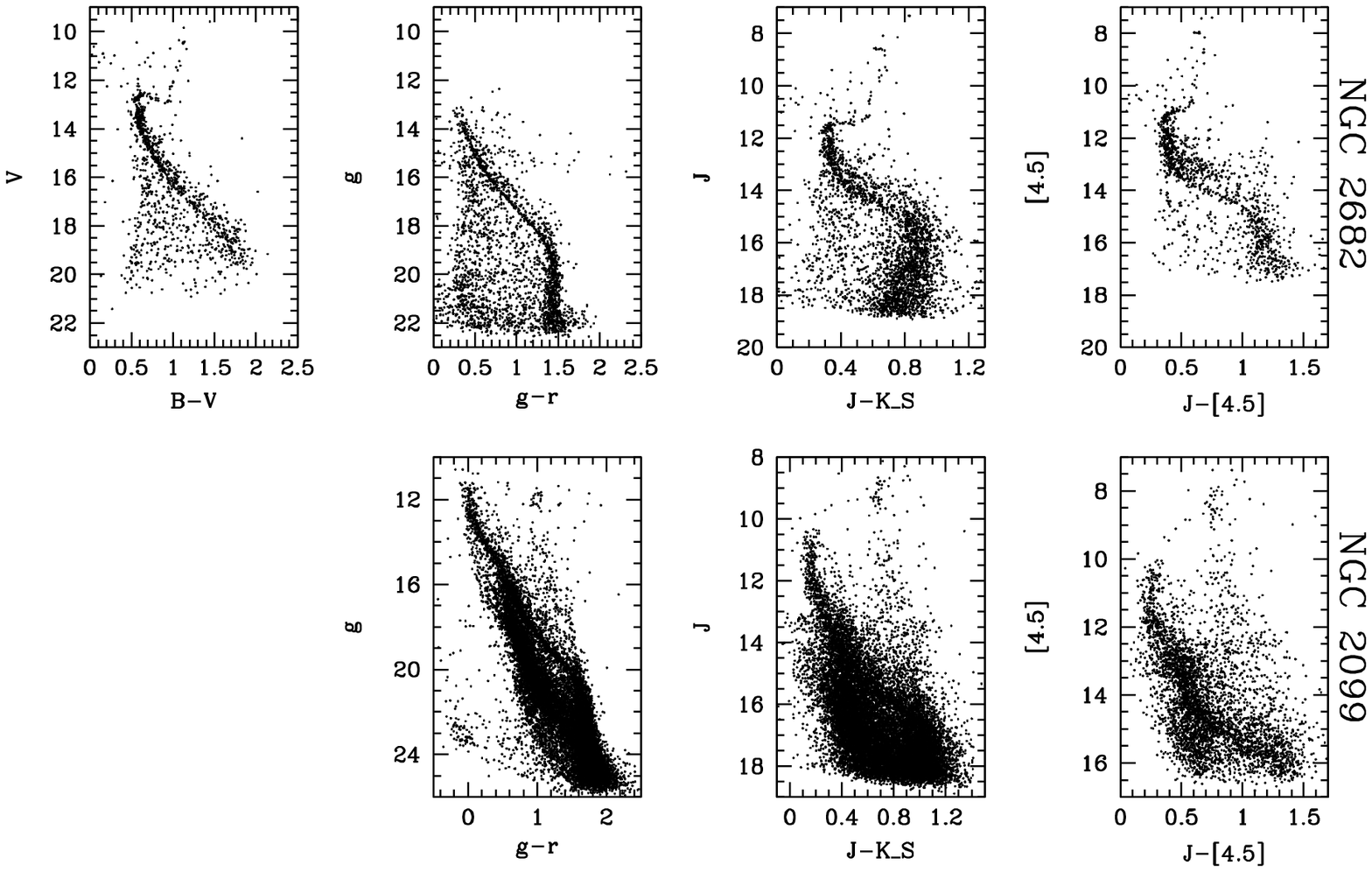}
\caption{Cluster CMDs for merged datasets. NGC 2682 includes supplementary $BVI$ photometry to include stars above the turn-off. CMDs are shown only for sources within 20$^\prime$ of the cluster centers. \label{fig:finalphot}}
\end{figure*}

\begin{figure*} \centering
\epsscale{1.1}
\plotone{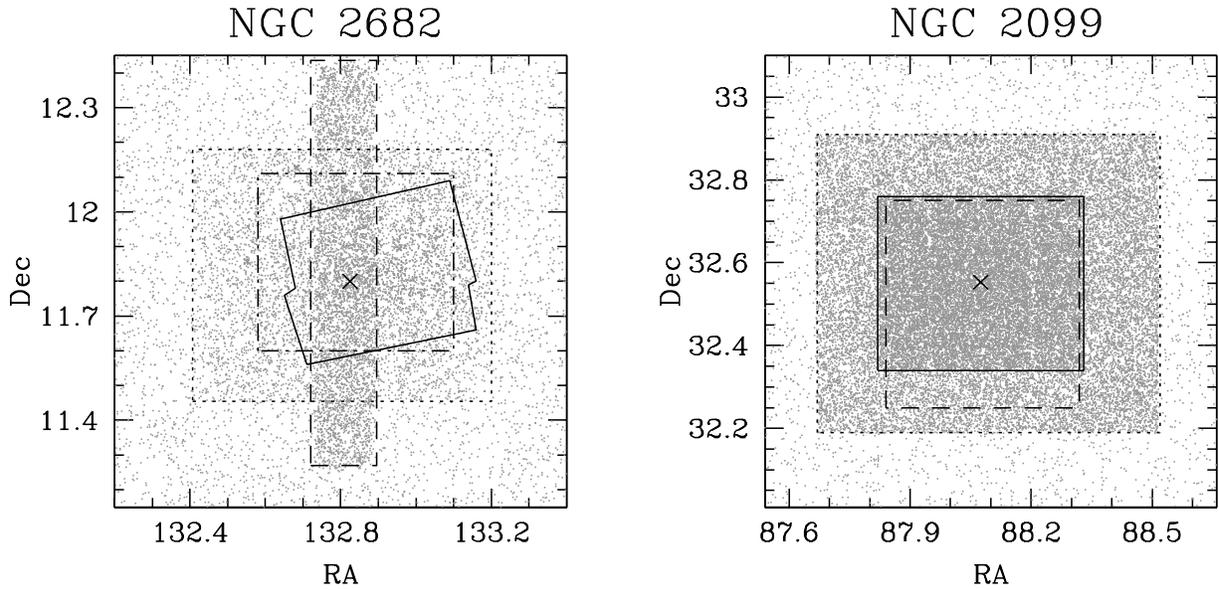}
\caption{Cluster spatial diagrams for merged datasets. \emph{Solid lines:} $gri$ photometry datasets: \citet{2008ApJ...675.1233H} for NGC 2099, \citet{2009ApJ...700..523A} for NGC 2682. \emph{Dotted lines:} NEWFIRM $JHK_S$ photometry. \emph{Dashed lines:} IRAC mid-IR photometry. \emph{Dot-dash lines:} Supplemental $BVI$ photometry from \citet{2008AA...484..609Y} for NGC 2682. 2MASS near-IR and WISE mid-IR photometry is available for all-sky. NGC 2682 SDSS DR7 photometry is available over the entire plotted area.\label{fig:spatial}}
\end{figure*}

Optical, near- and mid-IR photometry sets for each cluster are merged into a final dataset. Final cluster color-magnitude diagrams (CMDs) and spatial distributions are shown in Figures \ref{fig:finalphot} and \ref{fig:spatial}.

The IRAC coverage area in NGC 2682 is only a thin stripe in declination, designed to overlap the 2MASS 6x calibration area. WISE photometry is necessary to implement the \textsc{binocs} detection method across the entire cluster area.

\begin{deluxetable*}{lllDDDDDDD}
\tabletypesize{\ssmall}
\tablecaption{New NEWFIRM and IRAC photometry. \label{tab:NGC 2682stub}}
\tablewidth{0pt}
\tablehead{ \colhead{Cluster} & \colhead{RA} & \colhead{Dec} & \multicolumn{2}{c}{$J$} & 
\multicolumn{2}{c}{$H$} & \multicolumn{2}{c}{$K_S$} &
\multicolumn{2}{c}{[3.6]} & 
 \multicolumn{2}{c}{[4.5]} & 
\multicolumn{2}{c}{[5.8]} &
\multicolumn{2}{c}{[8.0]} 
\\[-2ex]
\colhead{(NGC)} &\colhead{(2000.0)} & \colhead{(2000.0)} & \multicolumn{2}{c}{(mag)} & 
\multicolumn{2}{c}{(mag)} & 
\multicolumn{2}{c}{(mag)} & 
\multicolumn{2}{c}{(mag)} & 
\multicolumn{2}{c}{(mag)} & 
\multicolumn{2}{c}{(mag)} & 
\multicolumn{2}{c}{(mag)} 
}
\decimalcolnumbers
\startdata
2682 & 132.73465 & 11.77622 & 14.189$\pm$0.011 & 13.613$\pm$0.013 & 13.445$\pm$0.016 & 13.410$\pm$0.024 & 13.451$\pm$0.069 & 13.415$\pm$0.040 & 13.402$\pm$0.083 \\
2682 & 132.73955 & 11.58258 & 13.921$\pm$0.011 & 13.410$\pm$0.011 & 13.297$\pm$0.017 & 13.208$\pm$0.020 & 13.247$\pm$0.056 & 13.181$\pm$0.035 & 13.205$\pm$0.067 \\
2682 & 132.74037 & 12.01129 & 14.141$\pm$0.012 & 13.702$\pm$0.011 & 13.671$\pm$0.022 & 13.557$\pm$0.023 & 13.618$\pm$0.051 & 13.536$\pm$0.036 & 13.608$\pm$0.076 \\
2682 & 132.74144 & 12.08310 & 14.283$\pm$0.012 & 13.744$\pm$0.012 & 13.624$\pm$0.021 & 13.566$\pm$0.022 & 13.551$\pm$0.055 & 13.488$\pm$0.037 & 13.573$\pm$0.070 \\
2682 & 132.74359 & 11.80183 & 14.073$\pm$0.011 & 13.635$\pm$0.011 & 13.556$\pm$0.015 & 13.486$\pm$0.023 & 13.524$\pm$0.059 & 13.489$\pm$0.033 & 13.367$\pm$0.065 \\
2682 & 132.74379 & 11.81218 & 13.993$\pm$0.011 & 13.571$\pm$0.009 & 13.512$\pm$0.014 & 13.438$\pm$0.021 & 13.443$\pm$0.039 & 13.416$\pm$0.030 & 13.422$\pm$0.068 \\
2682 & 132.74596 & 11.80808 & 14.299$\pm$0.011 & 13.778$\pm$0.010 & 13.713$\pm$0.017 & 13.586$\pm$0.018 & 13.512$\pm$0.045 & 13.536$\pm$0.033 & 13.527$\pm$0.057 \\
2682 & 132.74651 & 11.97064 & 14.615$\pm$0.014 & 13.908$\pm$0.014 & 13.887$\pm$0.025 & 13.670$\pm$0.019 & 13.732$\pm$0.053 & 13.689$\pm$0.038 & 13.613$\pm$0.068 \\
2682 & 132.74748 & 12.16521 & 14.022$\pm$0.016 & 13.567$\pm$0.021 & 13.569$\pm$0.035 & 13.283$\pm$0.030 & 13.393$\pm$0.044 & 13.362$\pm$0.033 & 13.278$\pm$0.053 \\
2682 & 132.74765 & 11.91457 & 14.125$\pm$0.015 & 13.673$\pm$0.018 & 13.573$\pm$0.024 & 13.480$\pm$0.022 & 13.504$\pm$0.039 & 13.451$\pm$0.033 & 13.329$\pm$0.053 \\
\multicolumn{10}{c}{\nodata} 
\enddata
\tablecomments{This table is available in its entirety in machine-readable form in the online journal. A portion is shown here for guidance regarding its form and content.}
\end{deluxetable*}

\section{Stellar Isochrone Models} \label{sec:isochrones}
The \textsc{binocs} code uses synthetic SEDs from isochrones to determine best-fit masses for each star in the clusters. Therefore, the mass determination from the \textsc{binocs} code is only as accurate as the underlying isochrones themselves.

\begin{figure} \centering
\epsscale{1.25}
\plotone{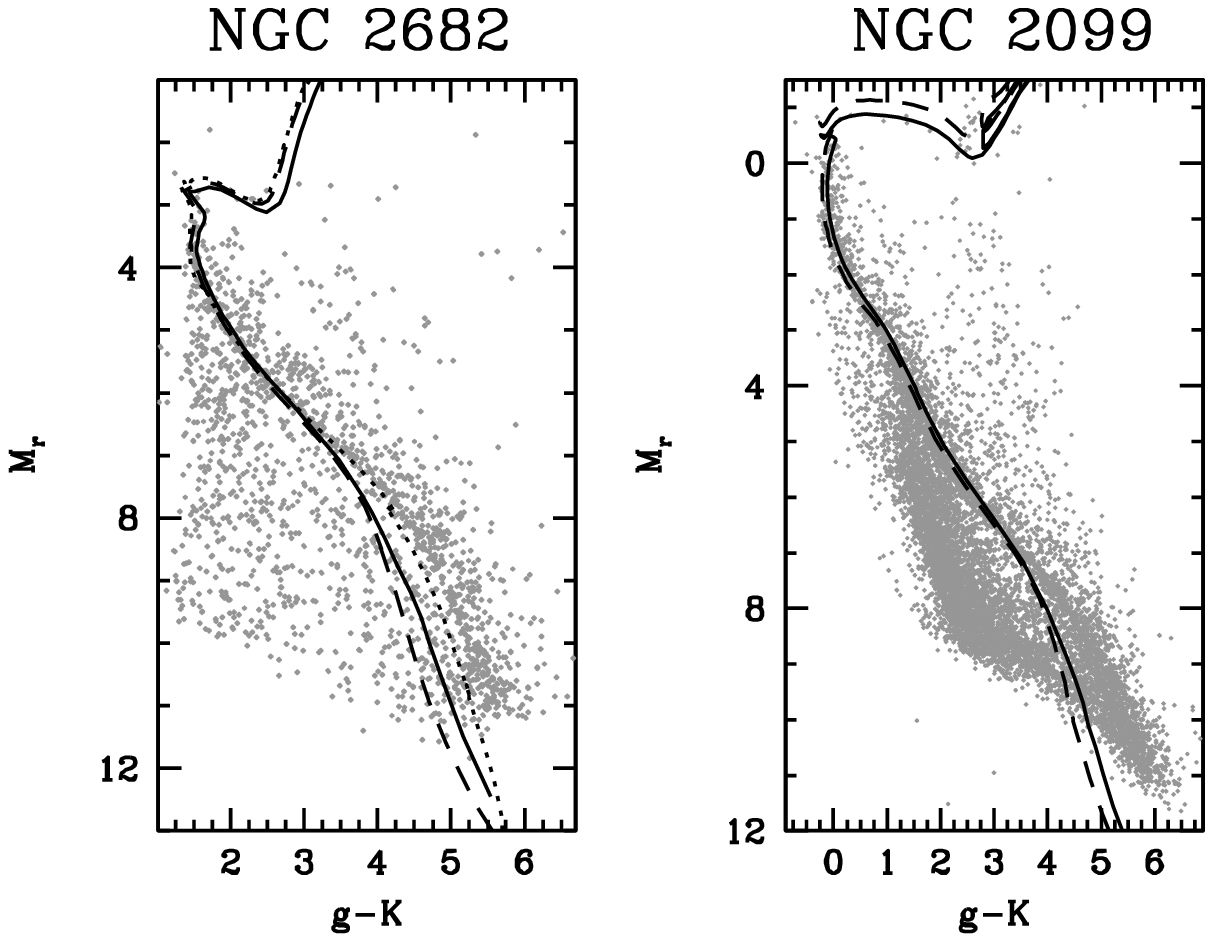}
\caption{Comparison of popular isochrone sets to combined cluster photometry in several CMDs. \emph{Dotted Line:} Dartmouth \citep[][NGC 2682 only]{2007AJ....134..376D}. \emph{Solid Line:} Padova \citep{2002AA...391..195G}. \emph{Dashed Line:} PARSEC \citep{2012MNRAS.427..127B}. NGC 2682 isochrones: 3.5 Gyr, [Fe/H] = +0.01, E($B-V$) = 0.04, 1386 pc. NGC 2099 isochrones: 355 Myr, [Fe/H] = +0.08, E($B-V$) = 0.3, 1386 pc. 
\label{fig:fidcmd}}
\end{figure}

Modern stellar models are still affected by non-negligible discrepancies due to variation in input physics \citep{2013AA...549A..50V}. This is apparent in the comparison of isochrone tracks computed by different sets of authors. In Figure \ref{fig:fidcmd}, two popular isochrone sets are over-plotted on NGC 2682 and NGC 2099 CMDs: Dartmouth \citep{2007AJ....134..376D} and  Padova \citep{2002AA...391..195G} or PARSEC \citep{2012MNRAS.427..127B}. For NGC 2682, 3.5 Gyr isochrones with [Fe/H] = +0.01, E($B-V$) = 0.04 and a distance of 855 pc were used. For NGC 2099, 355 Myr isochrones were used, with [Fe/H] = +0.08, E($B-V$) = 0.3 and a distance of 1386 pc. Dartmouth isochrones can only be interpolated in [Fe/H] for ages $> 1$ Gyr. Because no 355 Myr Dartmouth isochrones with [Fe/H] = +0.08 can be generated, they are not shown in Figure \ref{fig:fidcmd}.

\begin{figure*} \centering
\epsscale{1.1}
\plotone{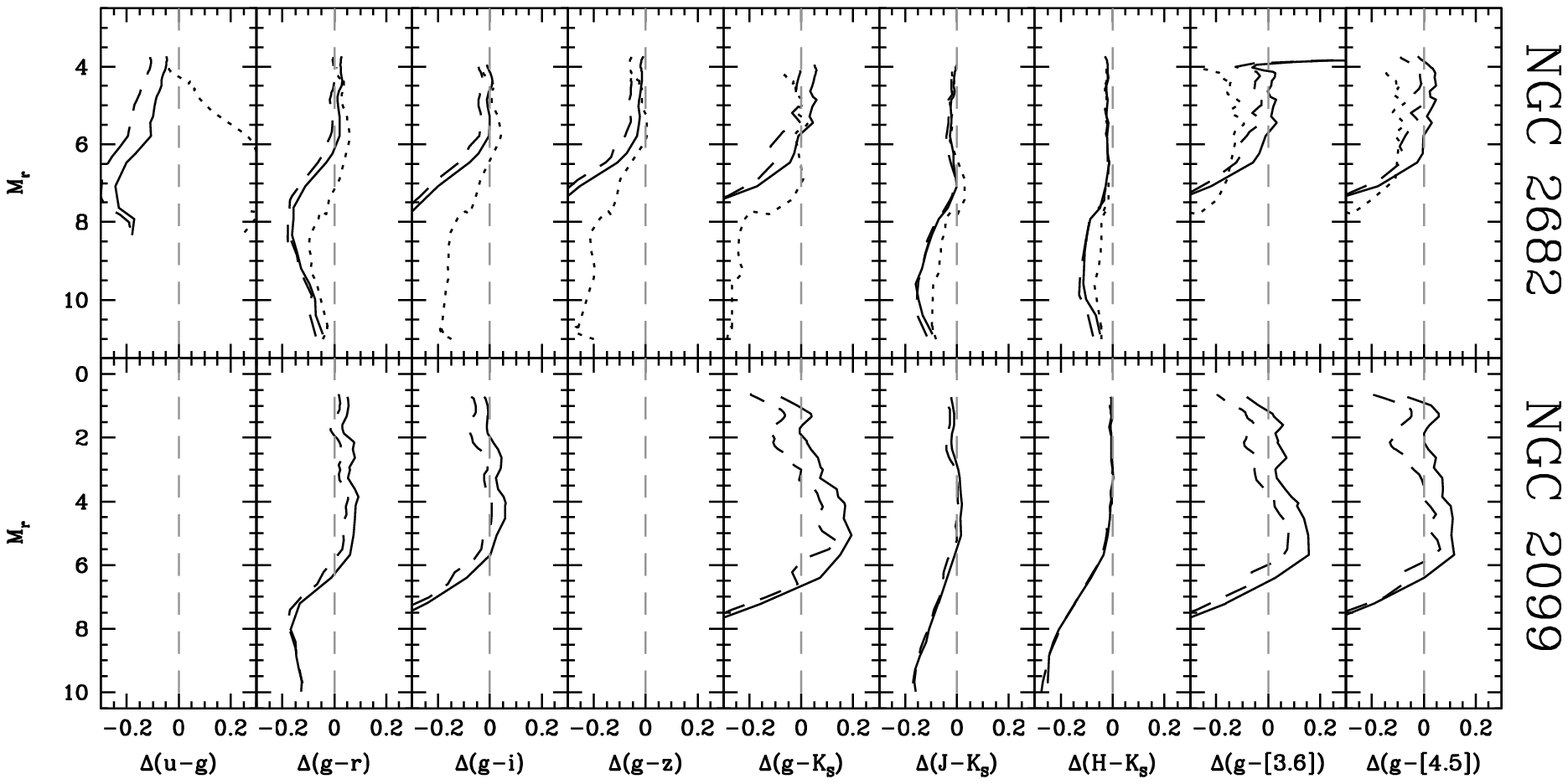}
\caption{Residuals between empirical ridgelines and isochrones for various filters. Same isochrones as shown in Figure \ref{fig:fidcmd}. \emph{Dotted Line:} Dartmouth \citep{2007AJ....134..376D}. \emph{Solid Line:} Padova \citep{2002AA...391..195G}. \emph{Dashed Line:} PARSEC \citep{2012MNRAS.427..127B}. \label{fig:fidrsd}}\vspace{-2mm}
\end{figure*}

It is clear from Figure \ref{fig:fidcmd} that all the isochrone sets deviate from the observed main sequence, especially for low-mass stars. To quantify the deviation between the models and observations, residuals between a by-eye empirical ridgeline and isochrones in various filters are shown in Figure \ref{fig:fidrsd}. 
The NGC 2682 $ugriz$ empirical ridgelines are pulled from the same source as the data, \citet{2009ApJ...700..523A}. For stars with magnitudes above or below the ridgeline area, colors are shifted such that the adjusted ridgeline is continuous.

Isochrone models vary anywhere from $\sim 0.1$ to over 0.3 in color, depending on filter. This large discrepancy in color may significantly affect results from the \textsc{binocs} fitting. Before being used to compute synthetic SEDs, these isochrones will have to be altered.

To correct the isochrones so that they more closely match the data, isochrones are adjusted to align with the hand-drawn ridgelines. 
This is done by assuming the bolometric correction to the $r$ filter (and hence the $r$ magnitude itself) is correct, and adjusting all colors accordingly to match the empirical ridgelines. This updated isochrone is then fed into the \textsc{binocs} code to create synthetic SEDs.

For all \textsc{binocs} runs, PARSEC isochrones were used. While the PARSEC isochrones showed the most deviation from the empirical ridgelines in Figure \ref{fig:fidrsd}, this error is corrected out by the empirical transformation. Of the isochrone sets considered, PARSEC provides the largest mass range of synthetic stars, and is therefore the most advantageous for this approach.

\section{\textsc{binocs} Testing} \label{sec:testing}
When the \textsc{binocs} code was introduced in \S\ref{sec:code}, several parameters were assumed: the number of iterations of the fitting, the number of ``good'' fitting filters, and the threshold to consider a magnitude ``good.'' Each of these parameters were tested, and the results are shown below.

\subsection{Number of Iterations}\label{sec:nruns}
The \textsc{binocs} fitting is iterated a number of times to produce best-fit masses and uncertainties. While the \textsc{binocs} code has random elements (sampling of Gaussian error distribution), if the process is iterated enough times, the final results will not vary greatly. Running excess iterations beyond this will use more computing time, but not enhance the results in any meaningful way. To determine the minimum number of iterations required, the combined NGC 2682 dataset was run through the \textsc{binocs} code with varying numbers of iterations: 3, 10, 30, 90, 150, 200, 300, 400, 500, 600, 700, 1200. 

For each number of iterations, the \textsc{binocs} code was run five times. Using these five runs, a ``\% uncertainty'', $\Sigma$, was computed for each star. $\Sigma$ is defined as the standard deviation of all five resulting masses divided by the average of the resulting masses for which the star is classified as a member. $\Sigma$'s for primary and secondary mass determinations are computed independently. Stars that were classified as non-member stars in all five runs (and hence not given any best-fit masses) were removed from the set.

\begin{figure} \centering
\epsscale{1.1}
\plotone{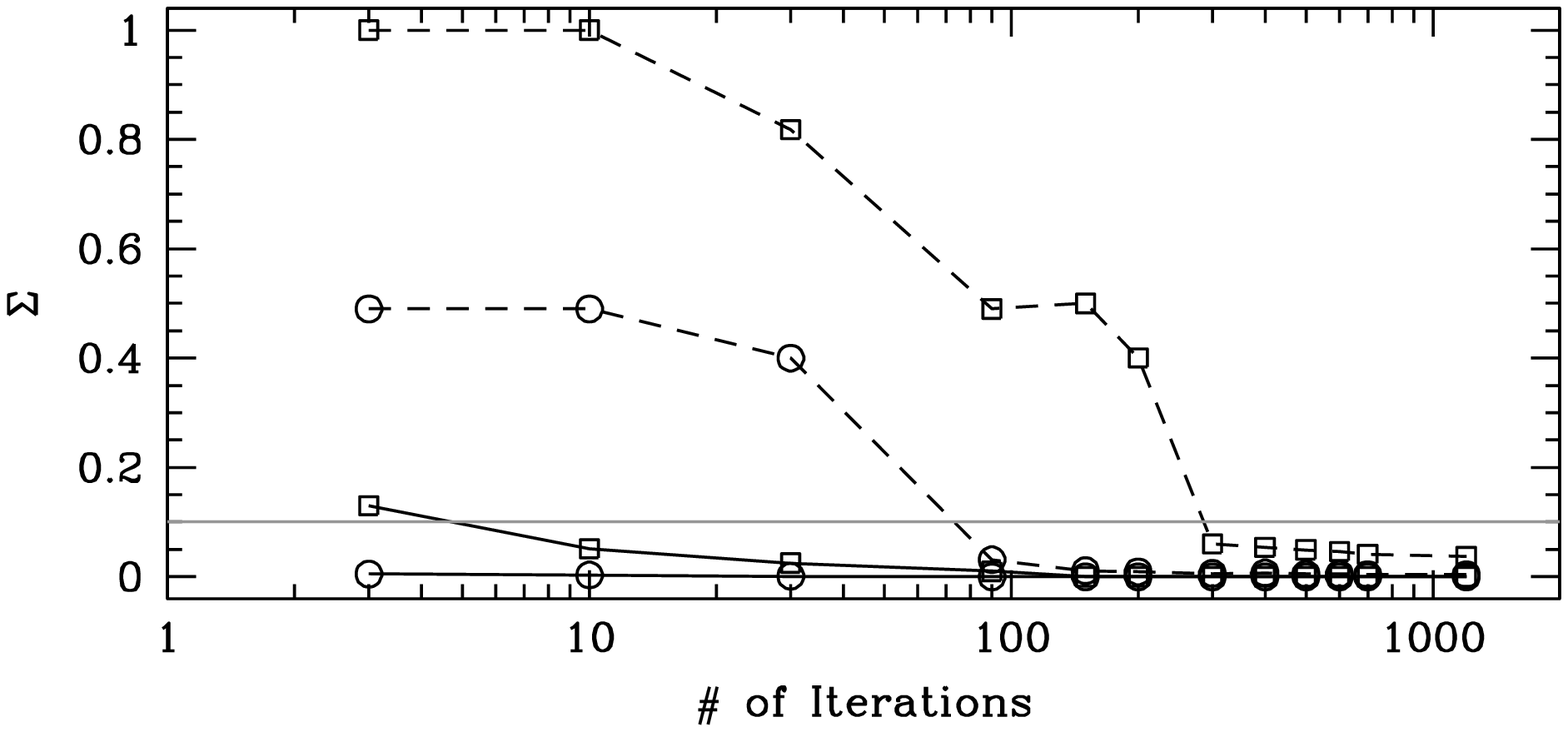}
\caption{Results of the number of iterations test. Circles correspond to $\Sigma$'s for primary masses, while squares correspond to secondary mass $\Sigma$'s. Solid lines show median $\Sigma$, dashed lines show 95$^{\text{th}}$ percentile. Grey line denotes average 10\% uncertainty between runs.\label{fig:nrun}}
\end{figure}

After computing $\Sigma$'s for all stars in the NGC 2682 dataset, median and 95$^{\text{th}}$ percentile $\Sigma$'s were computed for each iteration value. The results are shown in Figure \ref{fig:nrun}.

Median $\Sigma$ values are overall quite low; both primary and secondary median $\Sigma$'s are equal to zero for any number of iterations $\ge 150$. In order to ensure that a majority of stellar mass determinations are roughly constant between runs, we require the 95$^{\text{th}}$ percentile $\Sigma$ to be less than 0.1: on average, there will be a less than a 10\% difference in derived stellar masses between runs for 95\% of stars in the dataset. Using 300 iterations of the \textsc{binocs} fitting will satisfy this requirement (as seen in Figure \ref{fig:nrun}), and is chosen as the ideal number of iterations in the final computations.

\subsection{Number of Good Filters}\label{sec:ngoodfilters}

While a comprehensive sampling of the SED over all 10 filters ($UBVRI$ or $ugriz$, $JHK_S$, [3.6][4.5]) is ideal, it is often impractical to obtain quality photometry in this number of bands for every cluster we wish to study. In practice, the \textsc{binocs} code will have to produce accurate results using a less-than-ideal number of filters.

The library of synthetic SED models generated from the best-fit isochrone for NGC 2682 ([Fe/H] = +0.01, age = 3.55 Gyr) was used as an input into the \textsc{binocs} code. Using the combined NGC 2682 dataset, average photometric uncertainties were computed for all bands in bins of 0.5 mag. Every magnitude in the input library was randomized using a gaussian with $\sigma =$ 2 $\times$ the average photometric uncertainty in the corresponding bin.

The \textsc{binocs} code was run on the input library for various combinations of usable filters. For each run, only certain filter magnitudes in the randomized library were transferred to the final input file, listed in the first column of table \ref{tab:nfilters}.

Each filter combination was run 5 times, and each time the \% error in the primary and secondary mass determination was recorded. After all 5 runs, stars were binned into steps of 0.1 in mass ratio, and average \% error $+$ 1 standard deviation was computed for all stars in the bin. This 1$\sigma$ \% error is shown for each bin and filter combination in table \ref{tab:nfilters}.

\begin{table*} \centering
\caption{1$\sigma$ \% errors in mass estimates for various combinations of filters. \label{tab:nfilters}}
\begin{tabular}{lcccccccccccc|c}
 \hline \hline
\multicolumn{2}{c}{} & \multicolumn{11}{c}{Mass Ratio} & \multicolumn{1}{c}{} \\
\multicolumn{1}{c}{Filters} & $\phantom{jklm}$& 0.0 & 0.1 & 0.2 & 0.3 & 0.4 & 0.5 & 0.6 & 0.7 & 0.8 & 0.9 & \multicolumn{1}{c}{1.0} & \multicolumn{1}{c}{} \\ \hline 
101: $.g......$[3.6]. &&  6.6 &  0.0 &  0.1 &  1.1 &  1.2 &  2.5 &  4.1 &  4.9 &  5.1 &  6.2 &  9.6 &  \multirow{9}{*}{\begin{turn}{-90}1$\sigma$ \% Error in $M_{\text{pri}}$ \end{turn}} \\
111: $.g...J..$[3.6]. &&  6.0 &  0.0 &  0.1 &  0.9 &  1.0 &  1.8 &  3.1 &  3.7 &  3.9 &  4.7 &  9.1 & \\
202: $.gr.....$[3.6][4.5] &&  2.3 &  0.0 &  0.0 &  0.7 &  0.8 &  1.5 &  3.3 &  3.6 &  3.7 &  4.5 &  7.6 & \\
211: $.gr..J..$[3.6]. &&  2.7 &  0.0 &  0.0 &  0.7 &  0.7 &  1.8 &  3.0 &  3.6 &  3.8 &  4.5 &  7.9 & \\
212: $.gr..J..$[3.6][4.5] &&  2.5 &  0.0 &  0.0 &  0.7 &  0.7 &  1.3 &  2.9 &  3.4 &  3.4 &  4.1 &  7.7 & \\
222: $.gr..J.K_S$[3.6][4.5] &&  2.6 &  0.0 &  0.0 &  0.6 &  0.7 &  1.2 &  2.2 &  3.0 &  3.3 &  3.9 &  6.9 & \\
322: $.gri.J.K_S$[3.6][4.5] &&  1.5 &  0.0 &  0.0 &  0.6 &  0.7 &  0.9 &  1.9 &  2.4 &  2.7 &  3.3 &  6.1 & \\
332: $.gri.JHK_S$[3.6][4.5] &&  1.4 &  0.0 &  0.0 &  0.5 &  0.6 &  0.9 &  1.9 &  2.5 &  2.8 &  3.3 &  6.2 & \\
532: $ugrizJHK_S$[3.6][4.5] &&  0.6 &  0.0 &  0.0 &  0.2 &  0.2 &  0.5 &  0.8 &  1.4 &  1.8 &  1.9 &  4.4 & \\ \hline 

101: $.g......$[3.6]. &&  ... &  0.0 & 66.5 & 37.3 & 26.1 & 18.7 & 17.1 & 17.3 & 14.5 & 12.1 & 13.0 &  \multirow{9}{*}{\begin{turn}{-90}1$\sigma$ \% Error in $M_{\text{sec}}$ \end{turn}} \\
111: $.g...J..$[3.6]. &&  ... &  0.0 & 43.7 & 29.4 & 20.7 & 16.1 & 11.3 & 10.4 &  9.9 &  8.8 & 11.2 & \\
202: $.gr.....$[3.6][4.5] &&  ... &  0.0 & 51.6 & 24.4 & 15.7 & 11.9 & 10.5 &  9.0 &  7.5 &  7.2 & 10.0 & \\
211: $.gr..J..$[3.6]. &&  ... &  0.0 & 39.9 & 32.8 & 18.5 & 12.6 &  9.7 & 10.5 &  8.0 &  7.3 & 10.3 & \\
212: $.gr..J..$[3.6][4.5] &&  ... &  0.0 & 44.9 & 23.5 & 15.8 & 11.4 &  8.9 &  8.3 &  6.9 &  6.4 & 10.1 & \\
222: $.gr..J.K_S$[3.6][4.5] &&  ... &  0.0 & 27.6 & 16.6 & 13.4 &  9.1 &  7.2 &  7.3 &  6.5 &  6.2 &  8.6 & \\
322: $.gri.J.K_S$[3.6][4.5] &&  ... &  0.0 & 38.1 & 18.8 & 11.1 &  7.1 &  5.7 &  5.8 &  5.0 &  4.8 &  7.7 & \\
332: $.gri.JHK_S$[3.6][4.5] &&  ... &  0.0 & 19.7 & 15.0 & 10.6 &  8.3 &  5.5 &  5.6 &  5.3 &  4.8 &  7.8 & \\
532: $ugrizJHK_S$[3.6][4.5] &&  ... &  0.0 & 37.7 & 13.5 & 10.4 &  6.0 &  3.3 &  3.8 &  3.7 &  2.9 &  5.3 & \\ \hline\\
\end{tabular}
\end{table*}

The `332' ($griJHK_S$[3.6][4.5]) filter combination is chosen as the preferred option for this work. Only requiring 3 optical filters increases the number of usable clusters while only marginally increasing the uncertainty in the final \textsc{binocs} results. For primary mass estimates the '332' combination produces the second-lowest 1$\sigma$ \% errors of all scenarios, and produces secondary mass estimates good to within 20\%. The only better scenario is the full photometry set (532), which is difficult to obtain for many open clusters, especially depth in the $u$ filter. 

Using these results, there are 1500 stars in NGC 2682 and 3500 stars in NGC 2099 which have the necessary number of filters for a good \textsc{binocs} fitting.

\subsection{``Close'' Filter Threshold}\label{sec:filterthresh}
In addition to generating accurate mass estimates for cluster stars, the \textsc{binocs} method can mark stars as non-members if they do not have the required number of ``close'' filters. Therefore, the threshold ($\frac{1}{\lvert m_{\text{star}} - m_{\text{model}} \rvert + \eta_{soft}} < X$) which defines whether a filter is ``close'' will adjust the level of field star contamination within the sample. Conversely, if the threshold is too stringent, many legitimate member stars may be discarded from the sample.

To test for the optimal threshold level, an input catalog was created similarly to that used in \S\ref{sec:ngoodfilters}. The input file to the \textsc{binocs} code contained three copies of the synthetic library created in \S\ref{sec:ngoodfilters}: one at the same distance as NGC 2682, one shifted a distance modulus of 0.8 nearer, and one shifted a distance modulus of 0.8 further than NGC 2682. As the magnitude difference between the single-star main sequence and equal-mass binary sequence is 0.753, there should be no degeneracies between the three copies of the input library.

The \textsc{binocs} code was run on the input file for various values of the threshold. After the run was complete, two numbers were computed: the percentage of member stars (from the copy of the library at NGC 2682's distance) that were classified as non-members, and the percentage of non-member stars (from the other two copies of the library) that were classified as members. The best-fit ``close'' threshold value is chosen such that the sum of these two values is at a minimum.

\begin{figure}
\epsscale{1.1}
\plotone{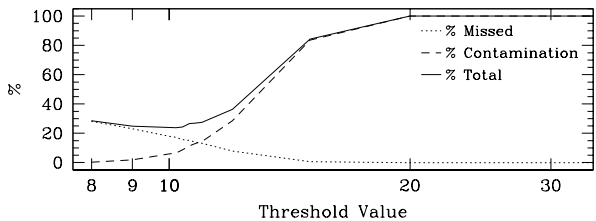}
\caption{Results of the threshold test for NGC 2682. Best threshold chosen to minimize \% Total.\label{fig:threshtest}}
\end{figure}

Figure \ref{fig:threshtest} shows the results of the threshold testing. A minimum in \% Total is found at a threshold value of 10. Contamination from foreground and background stars quickly increases when threshold values are larger than 10, while the percentage of missed member stars only decreases gradually.

A formal $\chi^2$ form of \eqref{eq:binocs} was also explored in addition to the fixed-width threshold form used:

\begin{equation}
\Theta = \sum_{\text{filters}} \frac{(m_\text{star} - m_\text{model})^2}{\sigma_\text{star}^2}
\end{equation}

\noindent While the $\chi^2$ form produced less theoretical contamination in this test, it also classified {\em more than 90\% of the stars in the NGC 2682 dataset as non-members!} As this is unrealistic for a cluster so far from the galactic plane, the fixed-width threshold form of \eqref{eq:binocs} was used in the final \textsc{binocs} code.

\subsection{Minimum Mass Ratio}\label{sec:minq}

It is often impossible to tell the difference between a single star and a low mass-ratio binary, even when using a minimum of 8 filters. A minimum mass ratio, as a function of primary mass, is determined to be the maximum of three values:

\begin{itemize}
\item \emph{Lowest mass-ratio model:} For the PARSEC isochrones being used, the lowest-mass model has a mass of 0.13 M$_\odot$, defining a minimum model mass ratio for each primary mass.
\item \emph{Synthetic best-fit mass ratio:} After each run of the \textsc{binocs} code, a test similar to that in \S\ref{sec:ngoodfilters} is run. Minimum mass ratios are defined as the resulting best-fit of each synthetic single star.
\item \emph{Constant threshold:} After a detailed comparison to clusters that have RV comparisons available, we find that \textsc{binocs} results prove unreliable, single/binary indistinguishable, for stars with $q < 0.3$ (see \S\ref{sec:rvcompare}).  Even if the synthetic tests estimate a value less than this, the minimum threshold for identifying a binary is a mass ratio  $q \ge 0.3$. 
\end{itemize}

\begin{figure}
\epsscale{1.1}
\plotone{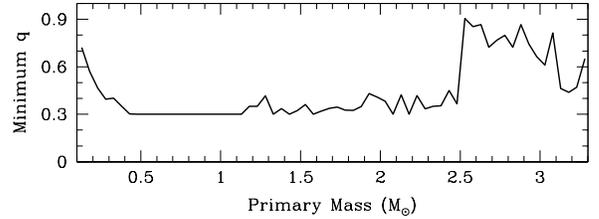}
\caption{Minimum mass ratios for NGC 2099, as a function of primary mass. \label{fig:minq}}
\end{figure}

Stars are defined to be singles if they have best-fit mass ratios below the specified value. Minimum mass ratios for NGC 2099 are shown in Figure \ref{fig:minq}. For systems with primary masses below 0.5 M$_\odot$, minimum mass ratios are dominated by the minimum model mass of 0.13 M$_\odot$. Above a primary mass of 2.5 M$_\odot$, minimum mass ratios are dominated by degeneracies at the turn-off.  As shown in Figure \ref{fig:minq}, the \textsc{binocs}  technique works well bewteen 2.5--0.5 M$_\odot$ for NGC 2099.

In their analysis, \citeauthor{2013ApJ...765....4D} stated their analysis was only sensitive to binaries with $q \ge 0.55$. The \textsc{binocs} method shows an improvement in mass sensitivity, with minimum mass ratios closer to 0.35 for a large mass range.

\section{Results}\label{sec:rvcompare}
The \textsc{binocs} results for NGC 2682 were compared to a published RV study of 104 cluster members (26 binaries) by \citet{1986AJ.....92.1364M}.
A comparison of multiplicity determinations is shown in table \ref{tab:rvcompare}.
The comparison was limited to stars with $14.5 \le g \le 16.5$, avoiding the degeneracies at the turn-off, and ending at the lower magnitude limit for the RV study.

\begin{figure}
\epsscale{1.22}
\plotone{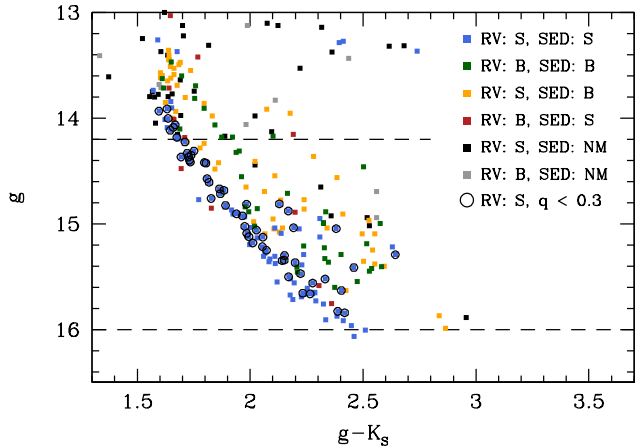}
\caption{CMD comparison of RV and \textsc{binocs} SED-fitting results for NGC 2682. Stars considered in comparison are those within dashed line limits. Color of dot indicates which cell of Table \ref{tab:rvcompare} star belongs to. Black circles indicate RV singles which were classified as \textsc{binocs} best-fit binaries with mass ratios $< 0.3$. \label{fig:rv_M67}}
\end{figure}

Using the updated minimum mass ratio calculation in \S\ref{sec:minq}, there is good agreement between \textsc{binocs} and RV results, with 60\% of RV singles being confirmed as single stars by the \textsc{binocs} routine, and 56\% of RV binaries being confirmed.

\begin{deluxetable}{lcccc}
\tablecaption{Comparison of \textsc{binocs} and \citet{1986AJ.....92.1364M} RV multiplicity results for NGC 2682. as shown in Figure \ref{fig:rv_M67}.\label{tab:rvcompare}}
\tablehead{
\colhead{} & \multicolumn{3}{c}{\textsc{binocs}} \\\cline{2-4} 
\colhead{} & \colhead{Single} & \colhead{Binary} & \colhead{Non-member} 
}
\startdata
RV Single & 81 & 29 & 25 \\
RV Binary & 4 & 24 & 15 
\enddata
\end{deluxetable}

Each method has its own limitations and biases, and exact agreement is not expected. RV surveys cannot detect long-period binaries, or those with high inclination, while \textsc{binocs} is insensitive to these parameters. These types of systems may account for many of the 29 RV-single stars that the \textsc{binocs} routine fit as binaries. The 4 RV-binary stars that \textsc{binocs} determined to be singles may be systems with small secondaries. RV shifts for small companions may still be appreciable, while the amount of contributed light to the SED is not. The RV and \textsc{binocs} methods are complementary techniques, but still show a large amount of overlap in the results.  Unfortunately neither NGC 2682 nor NGC 2099 have published double-line spectroscopic binaries with masses determined.

  The \textsc{binocs} results for NGC 2168 were compared to a published RV study of the cluster in \citet{Geller:2010hn}. A comparison of multiplicity determinations is shown in Table \ref{tab:rvcompare}. To avoid complications from the turn-off, and poor faint data in the RV studies, the comparison is limited to a specific magnitude range. For NGC 2168, only stars with $14.2 \le V \le 16.5$ are considered.

or stars classified as binaries by \textsc{binocs} , many are also classified as binaries by RV detection methods, with a 69\% overlap in NGC 2168. The \textsc{binocs}  code shows a lower match when classifying RV singles\footnote{The term ``RV singles'' is used to denote a system which does not show an appreciable velocity shift.}, with only  59\% of RV singles being classified as singles by \textsc{binocs}  in NGC 2168. To ensure reasonable agreement between \textsc{binocs} and \citet{Geller:2010hn}, a floor of 0.3 was set in the minimum mass ratio calculation.

\begin{figure}
\epsscale{1.2}
\plotone{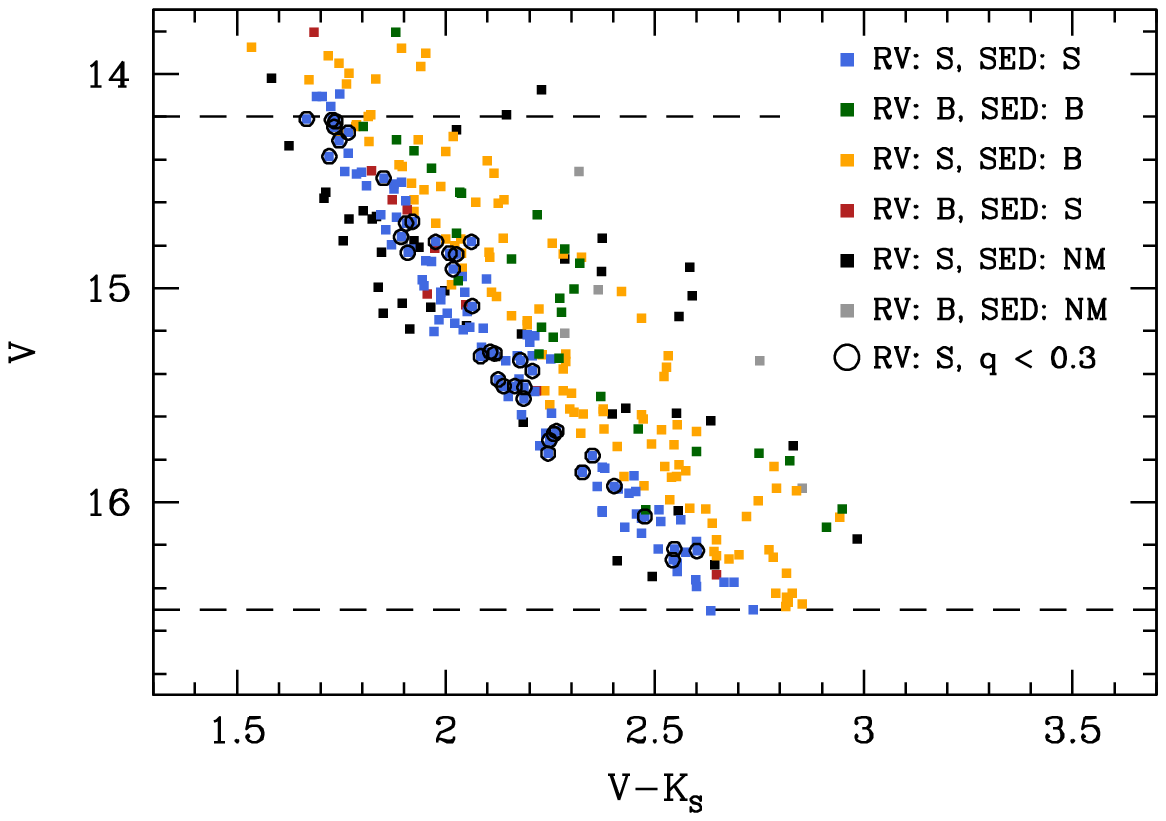}
\caption{CMD comparison of RV and \textsc{binocs}  SED-fitting results for NGC 2168. Stars considered in comparison are those within dashed line limits. Color of dot indicates which cell of table \ref{tab:rv_M35} star belongs to. Black circles indicate RV singles which were classified as \textsc{binocs}  best-fit binaries with mass ratios $< 0.3$. \label{fig:rv_M35}}
\end{figure}

\begin{deluxetable}{lcccc}
\tablecaption{Comparison of \textsc{binocs} and \citet{Geller:2010hn} RV multiplicity results, as shown in Figure \ref{fig:rv_M35}.\label{tab:rv_M35}}
\tablehead{
\colhead{} & \multicolumn{3}{c}{\textsc{binocs}} \\\cline{2-4} 
\colhead{} & \colhead{Single} & \colhead{Binary} & \colhead{Non-member} 
}
\startdata
Single & 113 (45\%) & 98 (39\%) & 40 (16\%) \\ 
Binary & 8 (18\%)  & 31 (69\%)  & 6 (13\%) 
\enddata
\end{deluxetable}

Using the updated minimum mass ratio calculation in \S\ref{sec:minq}, there is good agreement between \textsc{binocs} and RV results, with 60\% of RV singles being confirmed as single stars by the \textsc{binocs} routine, and 56\% of RV binaries being confirmed.

\vskip0.5in

\subsection{Mass Determination}

While not a part of the analysis set due to the lack of deep near-IR photometry, the clusters NGC 188 and NGC 6819 have also been the subject of detailed RV studies \citep[respectively]{2008AJ....135.2264G, 2009AJ....138..159H}. A comparison to the RV studies can be completed in the region where 2MASS photometry is available.

\begin{figure*} \centering
\epsscale{1.2}
\plotone{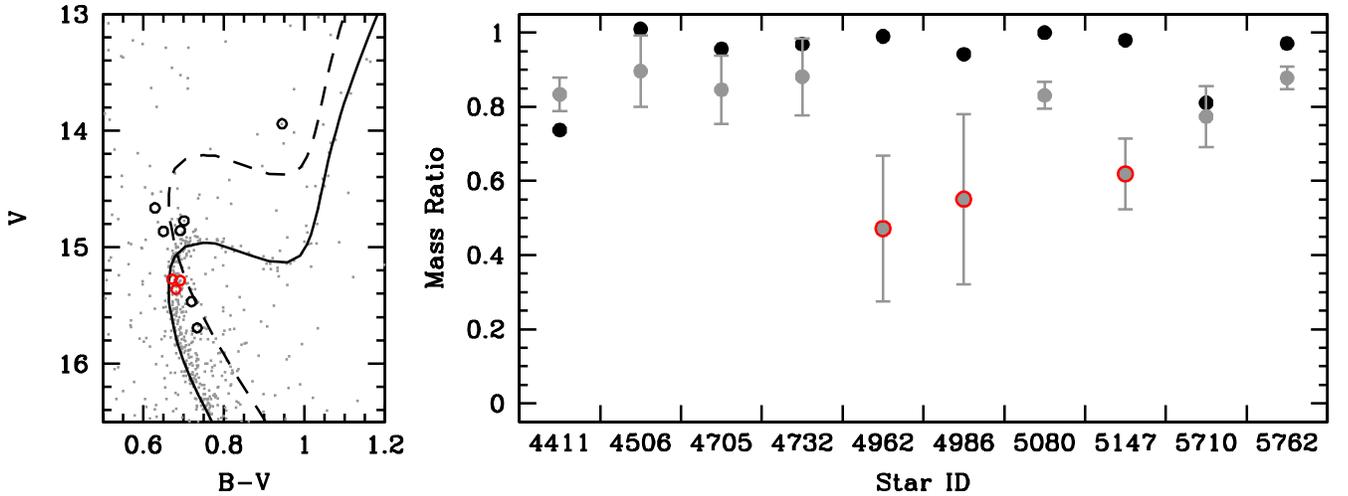}
\caption{\emph{Left:} NGC 188 CMD in $B-V$. Solid line is isochrone used to generate models for \textsc{binocs}  fitting. Dashed line is equal-mass binary sequence. Black circles are double-lined binaries. \emph{Right:} Comparison of RV mass ratios black) from \citet{2009AJ....137.3743G} to \textsc{binocs}  (grey) for NGC 188 double-lined binaries. Stars outlined in red are those complicated by degeneracies close to the turn-off.\label{fig:rv_n188}}
\end{figure*}

\begin{figure*} \centering
\epsscale{1.2}
\plotone{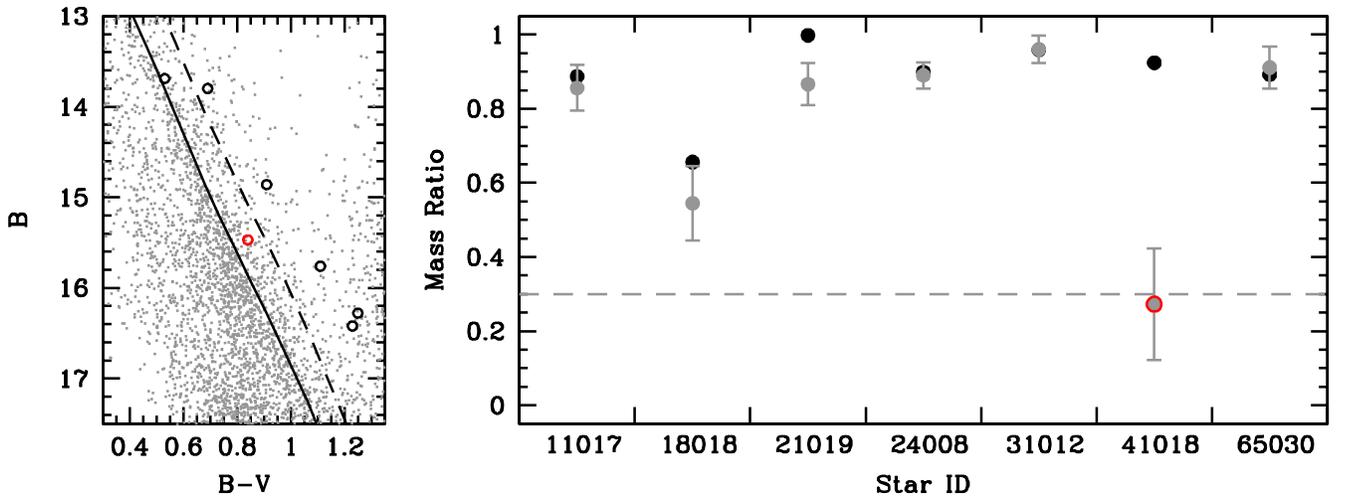}
\caption{Same as Figure \ref{fig:rv_n188} for NGC 2168. RV data from \citet{2015AJ....150...10L}. The stars outlined in red is below the $q \ge 0.3$ threshold level. \label{fig:rv_n2168}}
\end{figure*}

Of the 1046 stars studied in NGC 188, 13 were \emph{double-lined} binaries. Further follow-up on these stars, published in \citet{2009AJ....137.3743G}, characterized the orbits of these double-lined binaries, and produced accurate binary mass ratios. 
Similarly, NGC 6819 stars were followed up in \citet{2014AJ....148...38M}, and 15 double-lined binaries were detected. The RV-determined mass ratios are compared to those from \textsc{binocs}  in Figures \ref{fig:rv_n188} and \ref{fig:rv_n2168}.


There are several highly discrepant mass ratio determinations in NGC 188 and NGC 6819. Many of these double-lined systems lie near the turn-off of each cluster, where the single star main sequence and equal mass binary sequence overlap (as seen in the left hand panels of Figures \ref{fig:rv_n188}. In these regions, there are natural degeneracies, and the \textsc{binocs}  code cannot accurately determine parameters. Stars marked by red circles in Figures \ref{fig:rv_n188} lie extremely close to these degeneracies and therefore exhibit large errors with respect to the RV results.

Ignoring those stars very close to the crossing of single star main sequence and equal mass binary sequence, there is close agreement between RV and \textsc{binocs}  mass ratios. Including the quoted uncertainties in mass from \textsc{binocs}  (uncertainties from the RV surveys are negligible), mass ratios largely agree to within 10\%. \\

Combining this 10\% mass ratio accuracy with the previous conclusion that \textsc{binocs}  results are largely agreeing with RV multiplicity determinations, it is clear that the \textsc{binocs}  code is producing accurate results that can be extrapolated to lower-mass stars.

\subsection{Membership Comparison to {\em Gaia}}\label{sec:membership}

One additional effect of the \textsc{binocs} analysis is that stars are classified as single or binary members and non-members.  The ability to reject SEDs that cannot fit for stars of the correct distance, reddening, and luminosity class can be a powerful tool to exploring faint membership and binarity of simple populations.   

To test the effectiveness to photometric ``cleaning'' of non-member stars from the cluster CMD for main sequence stars, we have made a comparison to the {\em Gaia} DR2 \citep{gaia,gaiadr2} proper motion-based membership probability using the method from \citet{occam2} for two of our clusters NGC 2099 and NGC 2682. While the \textsc{binocs}  method will not be as effective as {\em Gaia}, the simplicity of using only photometry, allows probing much deeper than {\em Gaia}.

\begin{figure}
\epsscale{1.2}
\plotone{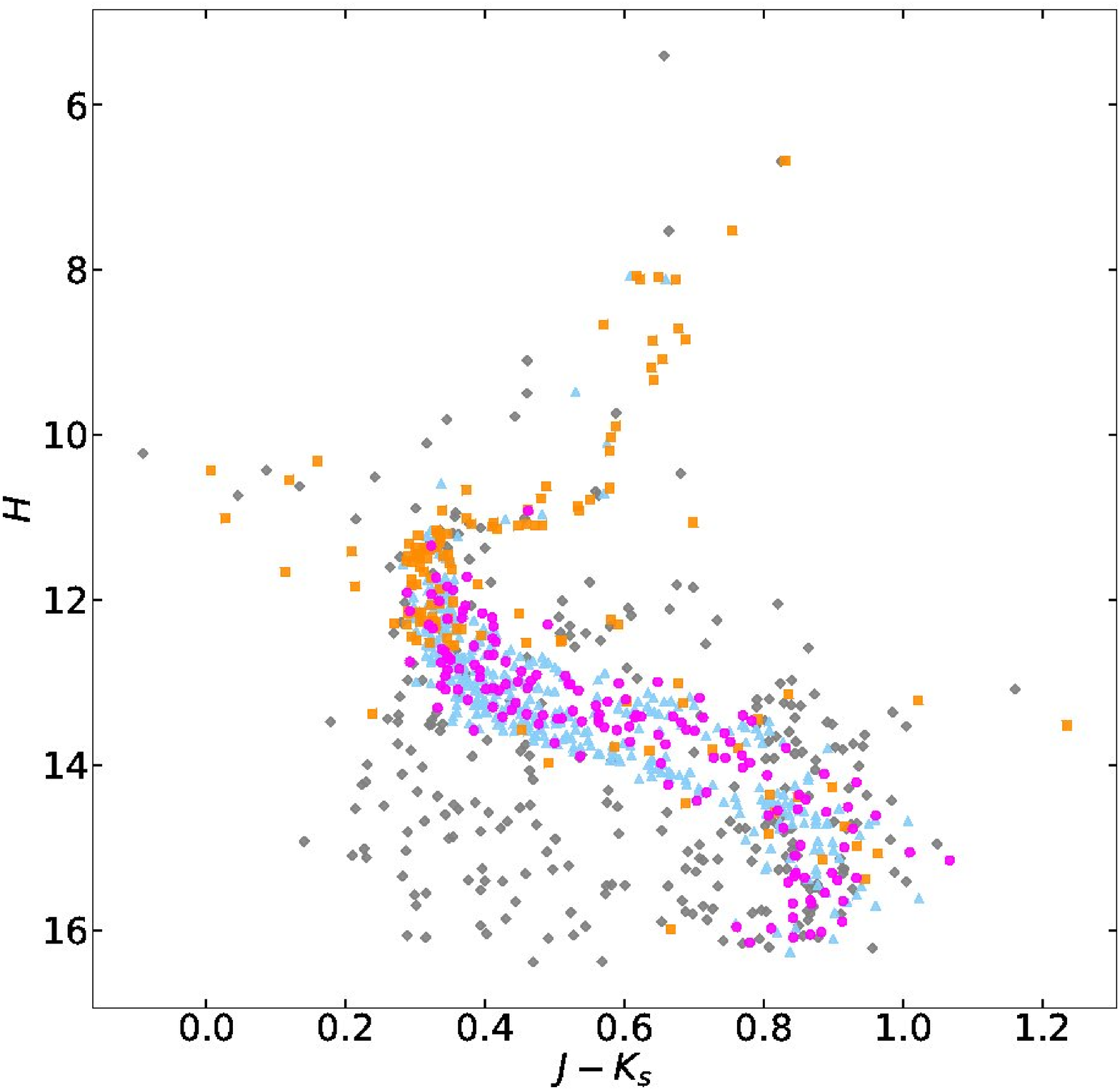}
\caption{2MASS CMD of cluster NGC 2682 stars analyzed by \textsc{binocs} with {\em Gaia} proper motion-based membership data. Grey diamonds represent stars where both methods agree that they are non-members. Cyan triangles represent stars that are considered members with both the {\em Gaia} method and the \textsc{binocs} method. Orange squares represent stars that are considered members with {\em Gaia} and non-members with \textsc{binocs}. Magenta circles represent stars that are considered non-members with {\em Gaia} and members with \textsc{binocs}.  Star counts in each category can be found in Table \ref{tab:membcompare}.  For $H \ge 12$ where the \textsc{binocs} method is effective, the two membership methods agree for 74\% of the stars. \label{fig:gaia2682}}
\end{figure}

\begin{figure}
\epsscale{1.2}
\plotone{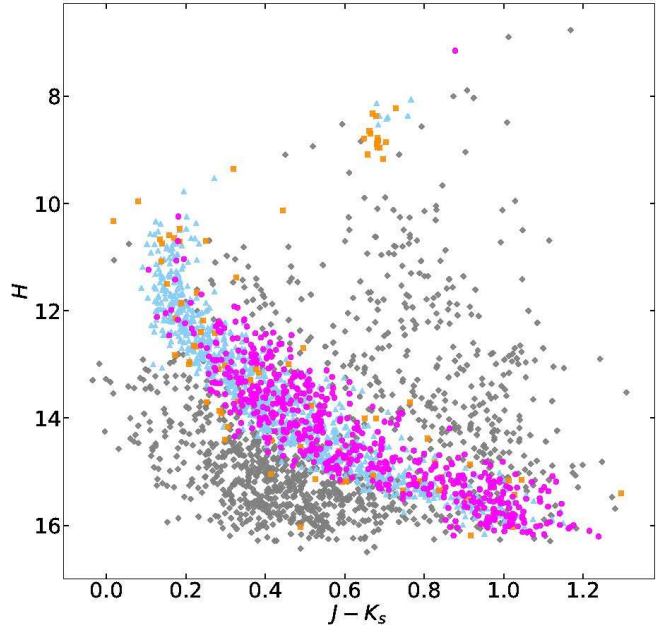}
\caption{Same as Figure \ref{fig:gaia2682}  for the cluster NGC 2099. These two methods agree for 75\% of the stars.\label{fig:gaia2099}}
\end{figure}

\begin{deluxetable}{llrr}[b]
\tablecaption{Comparison of  \textsc{binocs}  membership to {\em Gaia} membership\label{tab:membcompare}}
\tablehead{
\colhead{} && \colhead{\textsc{binocs}} & \colhead{\textsc{binocs}} \\[-3ex]
\colhead{} && \colhead{member} & \colhead{Non-member} 
}
\startdata
\hline & \multicolumn{2}{l}{NGC 2682 ($H \ge 12$)} \\[0.5ex]\hline
{\em Gaia} Member         & &    354  &  60    \\ 
{\em Gaia} non-member & &    138  & 229  \\
\hline & \multicolumn{2}{l}{NGC 2099 (all)} \\[0.5ex]\hline
{\em Gaia} Member         & &  886   & 82  \\ 
{\em Gaia} non-member & &   558  &  1152    \\
\enddata
\end{deluxetable}

We cross-matched the {\em Gaia} to the  \textsc{binocs} analyzed stars, where we then compared two methods to see how reliable they are in identifying clusters members: 1) the  \textsc{binocs}   membership method and 2) the Gaia proper motion-based membership method.   We found for the cluster NGC 2682 (Figure \ref{fig:gaia2682} and Table \ref{tab:membcompare}) that for stars fainter than $H \sim 12$, main sequence stars, the  \textsc{binocs}   method agrees fairly well with {\em Gaia} with \textsc{binocs} finding members for ($\sim$86\%) of the  {\em Gaia} members with significant overlap in members along the main sequence, which cuts-off at magnitudes $H \sim 16$ due to the limit
{\em Gaia} photometry and proper motions.   
However we do find that \textsc{binocs}  method is not quite as discriminating, as it finds 543 members compared to 414 with {\em Gaia}.

We preformed the same comparison for the cluster NGC 2099 (Figure \ref{fig:gaia2099} and Table \ref{tab:membcompare}). The  \textsc{binocs}  method members again overlap with many of the Gaia members in main sequence portion of the CMD and cut-offs at magnitudes around 16, due to the same 
limitations. In this plot, far more stars in the  \textsc{binocs}  sample is considered to be members of NGC 2099, as compared to NGC 2682, since this younger cluster has a longer main sequence where the  \textsc{binocs}  works well.   For this cluster, \textsc{binocs} categorizes $\sim$92\% of the {\em Gaia} members correctly as members, but again includes more {\em Gaia} non-member stars as members.

As a comparison to {\em Gaia} proper motion membership, one of the best membership methods available, we find  \textsc{binocs} agrees with {\em Gaia} membership $>75$\% of the time on the main sequence, but is more inclusive on non-members.  However,  \textsc{binocs}  can be used at large distances, unlike {\em Gaia}, such as to explore membership and binarity in simple stellar populations in other galaxies, given sufficient photometric depth (e.g., {\em Hubble + JWST}).

 \subsection{Binary Fractions}
  
After validating the \textsc{binocs}  code, it can begin to be applied to the clusters in the analysis set with the requisite photometry. The \textsc{binocs}  code was run on each of the 8 clusters available for this analysis (see Table \ref{tab:clusterParameters}), and the overall binary fraction was recorded. A list of clusters, their parameters, and the associated overall binary percentage is shown in Table \ref{tab:overallfbin}.

\begin{table*} \centering \scriptsize
\caption{Overall binary fractions for the 8 clusters considered in this analysis. \label{tab:overallfbin}}
\begin{tabular}{lccccrc} \hline\hline
				 & \textbf{Binary}	 & \textbf{Age}		& 					& \textbf{R$_{gc}$}	& \textbf{Number of}	& \textbf{Mass} \\
\textbf{Cluster} & \textbf{Fraction} & \textbf{(Gyr)}   & \textbf{[Fe/H]}	& \textbf{(kpc)} 	& \textbf{Members}		& \textbf{Range (M$_\odot$)} \\ \hline
NGC 188			 & 0.44 			 & 6.30  			& $-$0.02				& 9.54				& 405  					& 0.80 -- 1.14 \\
NGC 1960 (M36)	 & 0.66 			 & 0.03 			& \nodata  				& 9.81				& 941  					& 0.65 -- 6.46 \\
NGC 2099 (M37)	 & 0.48 			 & 0.35 			& $+$0.08				& 9.88				& 1632 					& 0.32 -- 3.21 \\
NGC 2158		 & 0.49 			 & 1.10  			& $-$0.23				& 13.56				& 266  					& 1.00 -- 1.98 \\
NGC 2168 (M35)	 & 0.61 			 & 0.13 			& $-$0.21				& 9.37				& 2258 					& 0.55 -- 3.19 \\
NGC 2420		 & 0.41 			 & 2.00  			& $-$0.23				& 10.81				& 748  					& 0.35 -- 1.63 \\
NGC 2682 (M67)	 & 0.44 			 & 3.50  			& $+$0.01				& 9.11				& 642  					& 0.19 -- 1.40 \\
NGC 6791		 & 0.39 			 & 8.00  			& $+$0.38				& 8.11				& 524  					& 0.89 -- 1.16 \\ \hline
\end{tabular}
\end{table*}

\subsubsection{Binary Fraction Versus Age}

\begin{figure}[h!]  \centering
\epsscale{1.2}
\plotone{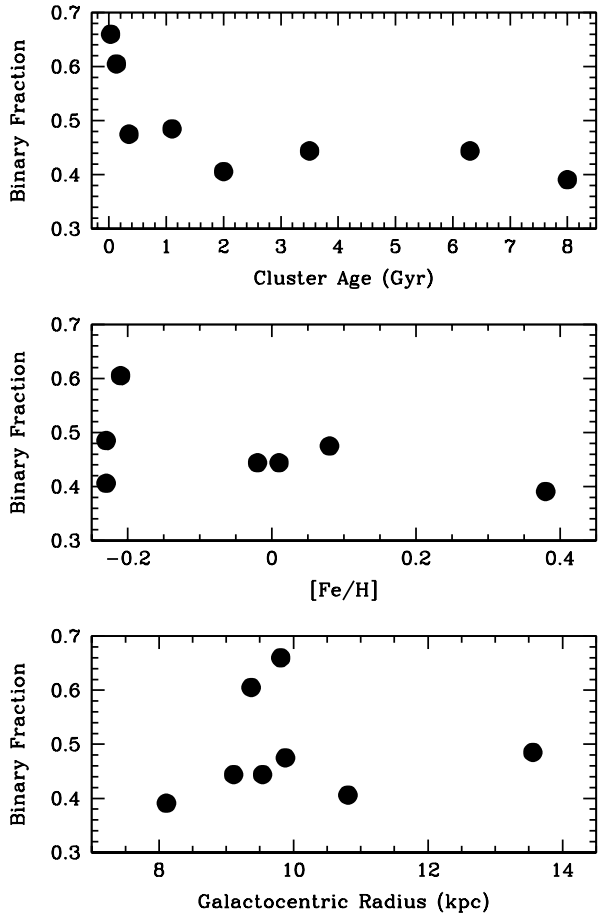}
\caption{(Top) Overall cluster binary fraction, as a function of cluster age. (Middle) verall cluster binary fraction, as a function of cluster [Fe/H]. (Bottom) Overall cluster binary fraction, as a function of cluster R$_{gc}$. \label{fig:fbinvage}}
\end{figure}

One of the main science questions of this work is how the binary population of a cluster evolves over time. The trend of overall binary fraction with cluster age is shown in Figure \ref{fig:fbinvage}. Overall, there seems to be a decreasing trend with age. Gravitational interactions between stars can easily disrupt some binary systems, while creating binaries from two single stars is much less common. It appears the majority of binary disruption occurs quickly during the first 200 Myr of a cluster's lifetime, after which the binary fraction becomes fairly constant.

After about 200 Myr, the binary fraction stabilizes to around 0.42, which is slightly higher than the measured binary percentage of 0.33 for field stars \citep{2009AAS...21333004R}. This small difference may be attributable to the fact that during the strong gravitational interaction which could eject a cluster binary system into the field population, the binary system may also be disrupted. Without a better understanding of the ejection processes of binary systems, the overall binary fraction of cluster and field stars cannot be easily compared.

Completing an analysis such as the one in Figure \ref{fig:fbinvage} using only RV surveys could take decades to build up enough analysis clusters to produce any useful insights. Two-band analysis, though fast, is dominated by degeneracies, and is limited to small magnitude ranges across the main sequence. With new, deep photometric surveys becoming available (UKIDSS, VVV, ESA {\it Gaia}, LSST), more clusters could be added to the list with minimal effort using the \textsc{binocs}  code. Generating the plot in Figure \ref{fig:fbinvage} using hundreds of open clusters would yield significant insights into the true distribution of binary fractions.

\subsubsection{Binary Fraction Versus Metallicity}

It is not well-understood how differences in metallicity of a pre-cluster cloud may affect the formation of binary systems. The distribution of binary fractions as a function of metallicity is shown in Figure \ref{fig:fbinvage}. There are only 7 clusters shown in Figure \ref{fig:fbinvage} due to the fact that M36 does not have a published metallicity value.

It is clear from Figure \ref{fig:fbinvage} that any observed trend will be dominated by the contribution from NGC 6791, at [Fe/H] $= +0.38$. Without this metallicity outlier, there is hardly any trend in binary fraction. The absence of a trend is still significant: the metallicity of a primordial cluster may have no effect on the binary population, at least on the aggregate level. This insight could be important for initial conditions of numerical simulations.

Similarly to the distribution with age, more data points can be added to this plot with minimal effort when new deep photometry becomes available. Filling in the remainder of the metallicity range will give more insight into whether a trend exists or not. Additionally, with a much larger number of clusters, binary fraction can be modeled as a function of both metallicity \emph{and} age. 

\subsubsection{Binary Fraction Versus Galactocentric Radius}

The above two comparisons have linked binary fraction to intrinsic cluster parameters, but clusters are not isolated systems, and the galactic environment plays a large part in cluster evolution. Clusters that are born near the center of the Galaxy experience a higher rate of tidal stripping events and other interactions which may alter the binary population. Figure \ref{fig:fbinvage} shows the overall binary fraction of clusters as a function of galactocentric radius (R$_{gc}$; the distance the cluster is from the center of the galaxy).

In Figure \ref{fig:fbinvage}, any observed trend is dominated by the two very young clusters, and thus high binary fraction, in the sample. Removing these two data points, a slight increasing trend with radius is observed. This would indicate that the gravitational shocking experienced at lower Galactic radii cause more binaries to be destroyed or ejected. However, NGC 2158, with a R$_{gc}$ of 13.5, is a high leverage point; its removal would result in there being no trend in R$_{gc}$. Additionally, the most central cluster is NGC 6791, with an age of 8 Gyr, while NGC 2158 has an age of 1.1 Gyr, an age difference which may explain the trend without R$_{gc}$.

As with the metallicity comparison, more clusters are needed to fill in the gaps in R$_{gc}$, disentangle correlations with age, and determine whether a trend truly exists. A more complete Figure \ref{fig:fbinvage} would allow the exploration of cluster-environment interactions, and would inform cluster simulations on the correct treatment of tidal stripping events and other gravitational collisions.

\section{Conclusions}
Understanding main sequence low-mass binary populations is essential for fully characterizing the masses and evolution of stellar clusters.  The characteristics of binary populations, such as the mass function and radial distribution are important for understanding the underlying physics of cluster evolution. It is well established that cluster stars, as well as high mass binary systems, undergo mass segregation over time, but the extent that this affects the low mass binary population has not been fully explored.   In this work: 
 
\begin{itemize}

\item We present new deep Near-IR and Mid-IR photometry for the open clusters NGC 2099 (M37) and NGC 2682 (M67).  The NOAO/NEWFIRM photometry reaches a depth of ($J$,$H$,$K_S$ = 18.6, 18.1, 17.8) for NGC 2099 and ($J$,$H$,$K_S$ = 18.8, 19.0, 18.0) for NGC 2682.  The {\em Spitzer}/IRAC photometry reaches a depth of ([3.6][4.5][5.8][8.0] = 18.0, 16.5, 14.6, 13.8 ) for NGC 2099 and ([3.6][4.5][5.8][8.0] = 18.5, 17.4, 15.0, 14.0) for NGC 2682.

\item We introduce the  \textsc{Binary INformation from Open Clusters using SEDs (binocs)}  a purely photometric method for determination of unresolved binaries and determination of the masses of both stars, for main sequence stars with primary masses below the turnoff to 0.5 M$_\odot$  (2.5--0.5 M$_\odot$ for NGC 2099).  We showed that the  \textsc{binocs}  method is a significant improvement over current binary detection techniques; requiring an order of magnitude less time, generating mass estimates on an order of magnitude more stars, and enabling quantitative exploration of faint binary systems, which are unreachable by RV studies.
  The  \textsc{binocs}  method allows for robust, quick binary classification that will become especially powerful as new all-sky surveys are released.

\item  We  tested the  \textsc{binocs}  code to ensure it produced reasonable results for binary detection and mass determination, when compared to previously-published studies based on multi-decade RV work.
Overall binary fractions can be computed quickly using  \textsc{binocs}   for clusters with sufficient photometry.

\item The results for NGC 188 are consistent with the result of \citet{Cohen20}, which compared to the \textsc{binocs}  results as preliminarily presented in \citet{Thompson15}. 
  
\item We find a clear decrease in binary fraction with respect to cluster age, due likely to disruption of wide-binary systems in the cluster environment.

\end{itemize}

\acknowledgements
The authors would like to thanks the referee for their patience and comments/suggestions that improved this paper. 

The authors would like to acknowledge graduate thesis travel support
from NOAO as well as financial support from the Texas Space Grant
Consortium and NSF-AST grant 1311835 and AST-1715662. Spitzer Cycle 3 observations
were funded under NASA/JPL sub-award grant GO-30800. The authors would
especially like to thank the GLIMPSE team, specifically Brian Babbler
\& Marilyn Meade, for their contribution in modifying the GLIMPSE
pipeline to handle our HDR IRAC data. The authors would also like to
thank the Max-Planck-Institut f\"ur Astronomie (MPIA Heidelberg) for hosting PMF and JD during the completion of this work.

This research uses services or data provided by the NOAO Science Archive. NOAO is operated by the Association of Universities for Research in Astronomy (AURA), Inc. under a cooperative agreement with the National Science Foundation. This work has also made use of the NASA/IPAC Infrared Science Archive, which is operated by the Jet Propulsion Laboratory, California Institute of Technology, under contract with the National Aeronautics and Space Administration.

\vskip0.1in

\bibliographystyle{aasjournal}
\bibliography{Binocs.ms}{}

\end{document}